\documentclass[aps,preprint,nofootinbib,superscriptaddress]{revtex4}%
\usepackage{amsmath}
\usepackage{autobreak}
\usepackage{amssymb}
\usepackage{amsfonts}
\usepackage{relsize}
\usepackage{enumerate}
\usepackage{booktabs}
\usepackage{color}
\usepackage{float}
\usepackage{placeins}
\usepackage{graphicx}
\usepackage{epstopdf}
\usepackage[caption=false]{subfig}
\usepackage{hyperref}
\usepackage[toc,page]{appendix}
\providecommand{\U}[1]{\protect\rule{.1in}{.1in}}
\providecommand{\U}[1]{\protect\rule{.1in}{.1in}}
\graphicspath{ {./image/} }
\providecommand{\U}[1]{\protect\rule{.1in}{.1in}}
\begin{document}
\title{Holographic Superconductors: An Analytic Method Revisit}
\author{En-Jui Chang}
\email{phyenjui@gmail.com}
\affiliation{Department of Electrophysics, National Chiao Tung University, Hsinchu, ROC}
\author{Chia-Jui Chou}
\email{agoodmanjerry.ep02g@nctu.edu.tw}
\affiliation{Department of Electrophysics, National Chiao Tung University, Hsinchu, ROC}
\author{Yi Yang}
\email{yiyang@mail.nctu.edu.tw}
\affiliation{Department of Electrophysics, National Chiao Tung University, Hsinchu, ROC}

\begin{abstract}
We study a non-minimal holographic superconductors model in both non-backreaction and full-backreaction cases using an analytic matching method. We calculate the condensate of the dilaton and the critical temperature of the phase transition. We also study the properties of the electric conductivity in various parameters.
\end{abstract}

\maketitle
\tableofcontents

\setcounter{equation}{0} \renewcommand{\theequation}{\arabic{section}.\arabic{equation}}

\section{Introduction}

The superconductivity is an extraordinary important phenomena in condensed matter physics.
It  was first discovered by Onnes  in 1911.
In 1933, a famous property of the superconductors, the Meissner effect, had been observed.
The symmetry breaking natural of the superconductor was then revealed by the phenomenological Ginzburg-Landau theory.
In 1957,  Bardeen, Cooper and Schrieffer proposed a microscopic theory which successfully describes the first type superconductivity, i.e. the BCS theory \cite{bardeen1957}.
However, the materials in the real world have plenty properties beyond the BCS theory.
The theoretical understanding of the strong-coupled superconductors still stayed in the barren land.

In the late 90's, Maldacena shed some light on the understanding of the notorious strong-coupled systems via the AdS/CFT conjecture, i.e. the holographic correspondence \cite{9711200,9802109,9802150,9905104,9905111}.
The holographic correspondence is a powerful tool to study a $d+1$-dimensional strongly coupled conformal field theories (CFT) by studying its $d+2$-dimensional dual gravitational theory in an Anti-de Sitter (AdS) space, and vice versa.
There has been an upsurge to study the strong-coupled superconductors via the holographic correspondence since the pioneer work by Steven S. Gubser \cite{0801.2977}, and this kind of the superconductors models are usually called the holographic superconductors (HSC) \cite{0803.3295,0810.1563,0904.1975,0906.1214,0912.0480,1002.1722}.

HSC has been widely studied during the last decade by using various numerical methods \cite{0810.1563,0908.3677,0912.0480,0912.3520,1006.2726,1007.3480,1309.0488,1410.7234,1607.08305,1704.00557,1804.05442,1805.07377,0810.1077,0906.2396,1003.3278,1101.3326,1207.3800,1604.08422,1605.07547,1607.08171,1608.04653,1609.05040,1609.08402,1610.07146,1610.08757,1611.06798,1611.07023,1612.04394,1612.06860,1612.07324,1703.00933,1706.05770,1706.07893,1707.05142,1708.01775,1710.07971,1710.10162,1712.04331,1801.06051,1801.06905,1803.05724,1803.06942,1803.08204,1804.06785,1003.0010,1608.00343,1705.08694,1708.01411,1710.05791,1710.07541,1711.07720,1808.07766,1005.1776,1105.5392,1202.0006,1302.4898,1404.2190,1305.6273,1411.6405} and analytic methods \cite{1601.04035,1606.04364,1608.05025,1708.04289,1710.09630,1712.07994,0907.3203,1103.5130,1105.4333,1107.2909,1211.0904,1307.5614,1603.02678,1609.09717,1702.07105,1705.10392,1708.06240}.
Many important properties of the strong-coupled superconductors have been successfully described by HSC.
Nevertheless, the analytical grasp is still naive, far from the exact solutions due to the extremely complicated non-linear equations of motion of the dual gravity theory.

An approximated analytic method called matching method was proposed in \cite{0907.3203} and has been used to study HSC recently \cite{0907.3203,1103.5130,1105.4333,1107.2909,1211.0904,1307.5614,1603.02678,1609.09717,1702.07105,1705.10392,1708.06240}.
In the matching method, the asymptotic expansions of the fields on the AdS boundary and the event horizon are matched at an intermediate point in the bulk.
So far the matching method  has only been applied in the non-backreaction case by treating the matter fields as probe fields.

In the non-backreaction case, the gravitational background is given and the probe matter fields are obtained by matching. 
Since the bulk behaviors and the gravity respondency of the fields in the bulk spacetime are sacrificed in the matching method, the matching solution is not the true solution of the equations of motion so that one might debate the validity of this method.
However, the matching method does describe the critical behaviors of the HSC very well both qualitatively and quantitatively.
We thus believe that the majority of the important physical properties of the holographic systems mainly depend on the asymptotic behaviors on the AdS boundary and the event horizon, but are not sensitive to the behaviors in the bulk at least in case of HSC.

Although the matching method applied in the non-backreaction case has successfully described the critical behaviors of HSC, the low temperature behavior is unsatisfactory. The condensation quickly becomes divergent when the temperature is cooled down from the critical value. This problem motivates us to study the HSC in the full backreaction case. In the full backreaction case, the gravitational background fields are obtained by matching as well as the probe matter fields. Unfortunately, there is an obstruction which makes it inapplicable to extend the matching method to the full-backreaction case directly because the differential equations of motion for the background fields are first-order equations. Nevertheless, since the key point of the matching method should be matching of the boundary conditions on the conformal boundary of the AdS spacetime and the event horizon, we generalize the matching method to collect all the degrees of freedom on the boundaries by relaxing their smooth conditions for the background fields. The details will be shown in section IV.

In this work, we consider a non-minimal HSC model with two parameters that has been studied numerically in \cite{1006.2726}. We study both the non-backreaction case using the ordinary matching method and the full-backreaction case using the generalized matching method.
We calculate the scalar condensation and the critical temperature. The generalized matching method in the full-backreaction case produces the consistent results  with the numerical ones in \cite{1006.2726}. In addition, we study the electric conductivity using an analytic truncating method.

This paper is organized as follows: 
In section II, we introduce the EMS system and describe the matching method to solve the system.
In section III, we study the condensate of dilaton in the non-backreaction case with the ordinary matching method.
In section IV, we study the condensate with the generalized matching method and the electric conductivity in the full backreaction case.
We summarize our results in section V.

(Unfortunately, there is a drawback of the matching method that the analytic description of the finite condensate region is still clueless.
An additional approximation is required to have an analytic behavior of the condensate \cite{1211.0904} at the temperature lower than the critical one.
Nevertheless, we found that the low temperature behavior is dominated by the backreaction of the matter fields on the metric of the bulk spacetime.
The $\chi$ field which appears in the metric only in the backreaction case plays a crucial role.

Is there any possibility to remove the approximations?
Our work is devoted to answer this question via generalizing the matching method and considering the full-backreaction.

There was an obstruction which makes the ordinary matching method inapplicable in the full-backreaction case.
A naive interpretation lingers in one's mind that the continuous and smooth matching condition might be necessary but not enough, although they are not physical conditions.
However, the main point of the matching method should be the matching of the boundary conditions on either the conformal boundary of the AdS spacetime or the event horizon in the bulk spacetime.
The generalized matching method collects all the boundary conditions by almost the same approach of the ordinary matching method.

The only difference between the generalized matching method and the ordinary one is relaxing the smooth matching condition for the matching of the gravity field.
The details will be shown later.)

\setcounter{equation}{0} \renewcommand{\theequation}{\arabic{section}.\arabic{equation}}

\section{Einstein-Maxwell-Scalar System}

In this paper, we consider a 4-dimensional Einstein-Maxwell-Scalar (EMS) system, which includes a gravity field $g_{\mu\nu}$, a Maxwell field $A_{\mu}$ and a charged complex scalar field $\Phi=\phi e^{i\theta}$.
After fixing the St\"{u}ckelberg field $\theta=0$, the action can be expressed as,
\begin{equation}
  S = \frac{1}{16\pi G_{4}} \int dzd^{3}x \sqrt{-g} \left[ R - \frac{f\left(\phi\right)}{4} F^{\mu\nu}F_{\mu\nu} + \frac{6}{\ell^{2}} U\left(\phi\right) - \frac{1}{2} \left(\partial^{\mu}\phi\partial_{\mu}\phi\right) - \frac{J\left(\phi\right)  }{2} A^{\mu}A_{\mu} \right],
  \label{action}
\end{equation}
where $f\left(\phi\right)$ is the gauge kinetic function which describes the interaction between the gauge field and the scalar field, $U\left(\phi\right)$ is the scalar potential and $J\left(\phi\right)$ is the extended St\"{u}ckelberg function that preserve the gauge invariant.
In this paper, we choose
\begin{subequations}
\begin{align}
  f\left(\phi\right)   &  = 1+\frac{\alpha^{2}}{2}\phi^{2},\\
  U\left(\phi\right)   &  = 1-\frac{\ell^{2}}{12}m^{2}\phi^{2},\\
  J\left(\phi\right)   &  = q^{2}\phi^{2}.
\end{align}
\end{subequations}
The similar system has been studied numerically in \cite{1006.2726}.
There are four free parameters $(\alpha,q,m,\ell)$ in the action.
We will analytically study the system by generalizing the matching method and compare our results with the ones in the numerical method.

The equations of motion are obtained by varying the action with the different fields,
\begin{subequations}
\begin{align}
  R_{\mu\nu} - \frac{1}{2}g_{\mu\nu} \left[ R - \frac{f\left(\phi\right)}{4} F^{\rho\sigma}F_{\rho\sigma} + \frac{6}{\ell^{2}} U\left(\phi\right) - \frac{1}{2} \left( \partial^{\rho}\phi\partial_{\rho}\phi \right) - \frac{J\left(\phi\right)}{2} A^{\rho}A_{\rho} \right]   &  \notag\\
  - \frac{f\left(\phi\right)}{2} F_{\mu\rho}F_{\nu}{}^{\rho} - \frac{1}{2}\partial_{\mu}\phi\partial_{\nu}\phi - \frac{J\left(\phi\right)}{2} A_{\mu}A_{\nu}  &  = 0,\\
  \nabla_{\mu} \left[ f\left(\phi\right)  F^{\mu\nu} \right] - J\left(\phi\right) A^{\nu}  &  = 0,\\
  \nabla^{2}\phi - \frac{1}{4} \frac{\partial f\left(\phi\right)}{\partial\phi} F^{\mu\nu}F_{\mu\nu} + \frac{6}{\ell^{2}} \frac{\partial U\left(\phi\right)}{\partial\phi} - \frac{1}{2} \frac{\partial J\left(\phi\right)}{\partial\phi} A^{\mu}A_{\mu}  &  = 0.
\end{align} \label{eom}%
\end{subequations}
Since we are going to study the thermodynamic properties in the HSC system at the finite temperature, we consider a black hole background that asymptotic to the AdS space. Without loss of generality, we consider the following ansatz of an isotropic black hole,
\begin{subequations}
 \label{ansatz}%
\begin{align}
  ds^{2}  &  = -g\left(r\right)  e^{-\chi\left(r\right)}dt^{2} + \frac{dr^{2}}{g\left(r\right)} + r^{2}\left( dx^{2}+dy^{2} \right),\label{metric}\\
  A  &  = A_{t}\left(r\right) dt, \\
  \phi &  = \phi\left(r\right),
\end{align}
\end{subequations}
where $r$ is the holographic radius which corresponds to the energy scale in the dual field theory.
The Hawking temperature of the black hole can be calculated as,
\begin{equation}
  T = \frac{1}{4\pi} g\left(r\right)^{\prime} \left. e^{-\frac{\chi(r)}{2}} \bigg|_{r=r_H} \right.,\label{fBT}
\end{equation}
where the black hole horizon $r_{H}$ is defined by $g(r_{H})=0$.
Plugging the ansatz (\ref{ansatz}) into the equations of motion (\ref{eom}) leads to the following equations of motion for the background fields:
\begin{subequations}
\begin{align}
  \chi^{\prime} + \frac{r}{2} \phi^{\prime2} + \frac{rJ}{2g^{2}} e^{\chi}A_{t}^{2}  &  = 0,\\
  \frac{g^{\prime}}{rg} + \frac{1}{r^{2}} + \frac{1}{4} \phi^{\prime2} - \frac{3}{\ell^{2}g} U\left(\phi\right) + \frac{f}{4g} e^{\chi}A_{t}^{\prime2} + \frac{J}{4g^{2}} e^{\chi}A_{t}^{2}  &  = 0,\\
  A_{t}^{\prime\prime} + \left( \frac{f^{\prime}}{f} + \frac{2}{r} + \frac{\chi^{\prime}}{2} \right) A_{t}^{\prime} - \frac{J}{gf}A_{t}  &  = 0,\\
  \phi^{\prime\prime} + \left( \frac{g^{\prime}}{g} + \frac{2}{r} - \frac{\chi^{\prime}}{2} \right) \phi^{\prime} + \left( \frac{6U^{\prime}}{\ell^{2}g} + \frac{e^{\chi}A_{t}^{\prime2}Jf^{\prime}}{2g} + \frac{e^{\chi}A_{t}^{2}J^{\prime}}{2g^{2}} \right) \frac{1}{\phi^{\prime}}  &  = 0,
\end{align} \label{eomr}%
\end{subequations}
where the prime represents the derivative with respect to $r$.
It is easy to verify that the RN-AdS black hole is a simple solution of the above equations of motion (\ref{eomr}),
\begin{subequations}
\label{RN}
\begin{align}
  g\left(r\right)  &  = \frac{r^{2}}{\ell^{2}} - \frac{1}{r} \left( \frac{r_H^3}{\ell^2} + \frac{\rho^2}{4r_H} \right) + \frac{\rho^2}{4r^2},\\
  A_{t}  &  = \rho \left( \frac{1}{r_H} - \frac{1}{r} \right),\\
  \chi\left(r\right)  &  = \phi\left(r\right) = 0.
\end{align}
\end{subequations}

In this paper, we are going to find an approximate analytic solution of the equations of motion (\ref{eomr}) using the matching method.
Instead of solving the equations of motion (\ref{eomr}) directly, we construct the asymptotic solutions first near both the horizon at $r=r_{H}$ and the boundary at $r=\infty$, respectively.
The solutions at the horizon and the boundary can be determined order by order given the appropriate boundary conditions.
We then match the two asymptotic solutions at some intermediate points $r_{m} \in (r_{H},\infty)$ by requiring the continuous and smooth conditions.
However, there is an obstruction which makes it inapplicable to extend the matching method to the full-backreaction case directly because the differential equations of motion for the background fields are first-order equations
We solve the problem  by relaxing the smooth matching conditions on the gravitational background fields.
The details of the generalized matching method will be described in the section IV.

Obviously, the solutions by matching are only good approximations at the horizon and the boundary, but not the exact solutions of the equations of motion at every $r$.
Nevertheless, since the core of the holographic correspondence is to relate the filed theory on the boundary and the near horizon geometry of the bulk, we believe that the detail behavior of the solution at the intermediate part does not affect the physics too much.
We thus can trust the physics from the results of this approximate matching solution at least qualitatively.

It is convenient to make a coordinate transformation from $r$ to $z=r_{H}/r\in(0,1)$ with the boundary at $z=0$ and the black hole horizon at $z=1$. 
In this work, to be concrete, we choose the matching point as $z_{m}=3/4$.
With the new coordinate $z$, the equations of motion for the background fields become,
\begin{subequations}
\label{eom-full}%
\begin{align}\label{g1}
  \chi^{\prime} - \frac{z}{2} \phi^{\prime2} - \frac{r_H^2B^2}{2z^3g^2}  &  = 0,\\
  g^{\prime} - \left( \frac{1}{z} + \frac{\chi^{\prime}}{2} \right) g - \frac{zf}{4} \left( B^{\prime} - \frac{\chi^{\prime}}{2} B \right)^{2} + \frac{3r_H^2}{z^3\ell^2} U  &  = 0,\\ \label{g2}
  B^{\prime\prime} + \left( \frac{f^{\prime}}{f} - \frac{\chi^{\prime}}{2} \right) B^{\prime} - \left( \frac{\chi^{\prime\prime}}{2} + \frac{\chi^{\prime}}{2} \frac{f^{\prime}}{f} + \frac{1}{z^4} \frac{r_H^2J}{gf} \right) B  &  = 0,\\
  \phi^{\prime\prime} + \left( \frac{g^{\prime}}{g} - \frac{\chi^{\prime}}{2} \right) \phi^{\prime} + \left[ \frac{\left(B^{\prime} - \frac{\chi^{\prime}}{2} B \right)^{2} f^{\prime}}{2g} + \frac{r_H^2B^2J^{\prime}}{2z^4g^2} + \frac{6r_H^2U^{\prime}}{z^4\ell^2g} \right] \frac{1}{\phi^{\prime}}  &  = 0.
\end{align}
\end{subequations}
where the prime represents the derivative with respect to $z$ now, and we have defined a new field $B^{2} = e^{\chi}A_{t}^{2}$ for convenience.

The series expansions of the background fields near the horizon can be written as: 
\begin{subequations}
\label{hts}%
\begin{align}
  \chi\left(z\right)  &  = \sum_{k=0}^{\infty}{\chi_{k}(1-z)^{k}}, \\
  g\left(z\right)  &  = \frac{r_H^2}{\ell^2 z^2} \left[1-z^{3} + \sum_{k=0}^{\infty}{g_{k}(1-z)^{k}}\right],\\
  B\left(z\right)  &  = \sum_{k=0}^{\infty}{B_{k}(1-z)^{k}}, \\
  \phi\left(z\right)  &  = \sum_{k=0}^{\infty}{\phi_{k}(1-z)^{k}}, 
\end{align}
\end{subequations}
In the above expansions, because the differential equations for $\chi$ and $g$ are first order and the differential equations for $B$ and $\phi$ are second order, there are six coefficients $(\chi_0, g_0, B_0, B_1, \phi_0, \phi_1)$ can be chosen as the boundary conditions at the horizon $z=1$, while the other higher ordered coefficients can be obtained from these six coefficients by solving the equations of motion order by order.

Specially, the boundary condition to define the horizon $g(1)=0$ leads $g_0=0$. Furthermore, the regularity boundary conditions at the horizon require,
\begin{subequations}
\begin{align}
  B_{0}  &  = 0,\\
  \phi_{1}  &  = -\frac{1}{g^{\prime}} \left( 6\frac{ \partial U\left(\phi\right) }{ \partial\phi } + \frac{B_1^2}{2} \frac{ \partial f\left(\phi\right) }{ \partial\phi } \right) \bigg|_{z=1}.
 \label{phi1}
\end{align}
\end{subequations}
In this work, for the least approximation, we only take the expansions at the horizon up to the order $k=1$ for the fields $\chi$ and $g$, and to the order $k=2$ for the fields $B$ and $\phi$. The series expansion of the background fields near the horizon become,
\begin{subequations}
\label{fHfull}%
\begin{align}
  \chi\left(z\right)  &  = \chi_{0} + \chi_{1}\left(1-z\right),\\
  g\left(z\right)  &  = \frac{{r_H}^2}{\ell^2 z^2} \left[ 1 - z^3 + g_1 \left(1-z\right) \right],\\
  B\left(z\right)  &  =  B_{1}\left(1-z\right) + B_{2}\left(1-z\right)^{2},\\
  \phi\left(z\right)  &  = \phi_{0} + \phi_{1}\left(1-z\right) + \phi_{2}\left(1-z\right)^{2},
\end{align}
\end{subequations}
where $\phi_1$ is given in Eq. (\ref{phi1}) and only three coefficients $(\chi_0, B_1, \phi_0,)$ left to be given.

At the boundary $z=0$, the asymptotic behavior of the fields are,
\begin{subequations}
\begin{align}
  \chi\left(z\right)  &  \sim \chi^{\left(0\right)} + \frac{\Delta_{\pm}}{4 \ell^2}\frac{D^2_{\Delta_{\pm}}}{{r_H^{2\Delta_{\pm}}}}z^{2\Delta_{\pm}},\label{chi}\\
  g\left(z\right)  &  \sim 
  \begin{cases}
    \frac{r_H^2}{\ell^2 z^2} - \frac{2M}{r_H}z + \frac{\Delta_{\pm}}{4 \ell^2}\frac{D_{\pm}^2}{{r_H^{2\Delta_{\pm}-2}}}z^{2\Delta_{\pm}-2} & \frac{1}{2} < \Delta_{\pm}\leq\frac{3}{2},\\
    \frac{r_H^2}{\ell^2 z^2} - \frac{2M}{r_H}z & \frac{3}{2} < \Delta_{\pm},
  \end{cases}\\
  B\left(z\right)  &  \sim \mu - \frac{\rho}{r_{H}}z,\\
  \phi\left(z\right)  &  \sim \frac{D_{\Delta_{-}}}{r_H^{\Delta_-}}z^{\Delta_-} + \frac{D_{\Delta_{+}}}{r_H^{\Delta_{+}}}z^{\Delta_+}
\end{align}
\end{subequations}
where
\begin{equation}
  \Delta_{\pm} = \frac{ 3\pm\sqrt{9+4m^2 \ell^2} }{ 2 }.
\end{equation}
are the conformal dimensions for the scalar fields with mass $m$. 

Similarly, in the above asymptotic forms, there are six coefficients $(\chi^{(0)}, M, \mu, \rho, D_{\Delta_{-}}, D_{\Delta_{+}})$ can be chosen as the boundary conditions at the boundary $z=0$. The asymptotic coefficients $\mu$ and $\rho$ in $B(z)$ are chemical potential and charge density. The asymptotic coefficients $D_{\Delta_{-}}$ and $D_{\Delta_{+}}$ in $\phi(z)$ represent the condensates. 

To satisfy the conformal symmetry, we should fix either $D_{\Delta_{-}}=0$ or $D_{\Delta_{+}}=0$ to obtain a stable solution.
Both boundary conditions are allowed and we choose $D_{\Delta_{-}}=0$ for convention in this work.
Setting $m^{2}\ell^2=-2$, we have $\Delta_{-}=1$ and $\Delta_{+}=2$.
With our choices, the asymptotic behavior of the background fields at the boundary $z=0$ becomes,
\begin{subequations}
\label{fBfull}%
\begin{align}
  \chi\left(z\right)  & x  \sim \chi^{(0)} +\frac{1}{2\ell^2}\frac{D_2^2}{{r_H^4}}z^4, \\
  g\left(z\right)  &  \sim \frac{r_H^2}{\ell^2 z^2} - \frac{2M}{r_H}z,\\
  B\left(z\right)  &  \sim \mu - \frac{\rho}{r_H}z, \\
  \phi\left(z\right)  &  \sim \frac{D_2}{{r_H^2}}z^2,
\end{align}
\end{subequations}
where only five coefficients $(\chi^{(0)}, M, \mu, \rho, D_{2})$ left to be given.

\setcounter{equation}{0} \renewcommand{\theequation}{\arabic{section}.\arabic{equation}}

\section{Non-Backreaction}

We first consider the case of non-backreaction, i.e. we treat the field $\phi$ and $B$ as the probe fields which do not affect the spacetime background. By setting $\phi=B=0$, the equations of motion (\ref{eom-full}) for the background fields reduces to
\begin{subequations}
\begin{align}
  \chi^{\prime}  &  = 0,\\
  g^{\prime} - \frac{g}{z} + \frac{3{r_H^2}}{z^3\ell^2}  &  = 0,
\end{align}
\end{subequations}
which admits the simple AdS-Schwarzschild black hole solution,
\begin{subequations}
\begin{align}
  \chi\left(z\right)  &  = 0,\\
  g\left(z\right)  &  = \frac{r_H^2}{\ell^2z^2} \left(1-z^3\right),
\end{align}
\end{subequations}
with the black hole temperature $T=3r_H^2/l^2$.

Next, we are going to solve the equations of motion for the probe fields $B$ and $\phi$,
\begin{subequations}
\label{eom-non}%
\begin{align}
  B^{\prime\prime} + \frac{f^{\prime}\left(z\right)}{f} B^{\prime} - \frac{1}{z^4} \frac{r^2_HJ}{gf} B  &  = 0,\\
  \phi^{\prime\prime} + \frac{g^{\prime}}{g} \phi^{\prime} + \left[ \frac{B^{\prime2}f^{\prime}}{2g} + \frac{{r_H^2}B^2J^{\prime}}{2z^4g^2} + \frac{6{r^2_H}U^{\prime}}{z^4\ell^2g} \right] \frac{1}{\phi^{\prime}}  &  = 0.
\end{align}
\end{subequations}
There are two scaling symmetries in the equations of motion (\ref{eom-non}) with the form $C \rightarrow s^{-n_{C}} C$, where $C$ is one of the variables $(r,t,\vec{x},g,B,\ell,q) $, $s$ is a scaling factor and $n_{C}$ is the scaling dimension for $C$.
The scaling dimensions for the two scaling symmetries are listed in Table \ref{scaling}.

\begin{table}[H]
\centering
\scalebox{1}{
\begin{tabular}
  [c]{ |c|c|c|c|c|c|c|c| } \hline
  Sym. & $n_{r}$ & $n_{t}$ & $n_{\vec{x}}$ & $n_{g}$ & $n_{B}$ & $n_{\ell}$ & $n_{q}$ \\ \hline
  I & 1 & -1 & -1 & 2 & 1 & 0 & 0 \\ \hline
  II & 1 & 0 & 1 & 0 & 0 & 1 & -1 \\ \hline
\end{tabular}}
\caption{The two scaling symmetries of the equations of motion (\ref{eom-non}) in the case of non-backreaction.
}\label{scaling}
\end{table}
We can use the above two scaling symmetries to set $\ell=1$ and $\rho=$ constant.
The generalized matching method is reduced to the ordinary matching method because there are only second order differential equations of motion.

In the case of non-backreaction, the series expansions of the fields $B$ and $\phi$ near the horizon in Eqs.(\ref{fHfull}) becomes,
\begin{subequations}
\label{fH}%
\begin{align}
  B\left(z\right)  &  =  a\left(1-z\right) + B_{2}\left(1-z\right)^{2},\\
  \phi\left(z\right)  &  = b+\phi_1 \left(1-z\right) + \phi_{2}\left(1-z\right)^{2},
\end{align}
\end{subequations}
where we have renamed $B_1=a$ and $\phi_0=b$ in (\ref{fHfull}). $a$ and $b$ will be imposed as the boundary values at the horizon $z=1$. Plug the series expansions (\ref{fH}) into the equations of motion (\ref{eom-non}),  $B_{2}$, $\phi_1$ and $\phi_{2}$ can be solved in terms of the boundary values  $a$ and $b$ order by order,
\begin{subequations}
\begin{align}
   B_{2}  &  = \frac{1}{6} \frac{  \big( \left( 4\alpha^2 + 2q^2 \right) r_H^2 + a^2\alpha^4 \big) b^2a  }{  r_H^2 \left( \alpha^2b^2 + 2 \right)  }, \\
  \phi_{1}  &  =  -\frac{1}{6}\frac{b \left( \alpha^2a^2 + 4r_H^2 \right)}{r_H^2} \\
    \phi_{2}  &  = -\frac{1}{48}\frac{  b  }{  r_H^4 \left( \alpha^2b^2 + 2 \right)  } \Bigg\{ \left( \frac{  32\alpha^2b^2  }{  3  } + \frac{64}{3} \right) r_H^4 \notag \\ 
  &  + 4a^2 \left[ -\frac{1}{3}\alpha^4b^2 + \left( b^2q^2 - \frac{10}{3} \right) \alpha^2 + \frac{2}{3}q^2 \right] r_H^2 + a^4\alpha^4 \left( \alpha^2b^2 - \frac{2}{3} \right) \Bigg\}.
\end{align}
\end{subequations}
At the boundary $z=0$, the asymptotic behavior of the fields $B$ and $\phi$ are the same as in Eqs. (\ref{fBfull}),
\begin{subequations}
\label{fB}%
\begin{align}
  B\left(z\right)  &  \sim \mu - \frac{\rho}{r_{H}}z,\\
  \phi\left(z\right)  &  \sim \frac{D_{2}}{r_H^{2}}z^{2}.
\end{align}
\end{subequations}
where $\mu$ and $D_2$ will be imposed as the boundary values at the boundary $z=0$.

The two boundary values $a$ and $b$ at the horizon $z=1$ and the two boundary values  $\mu$ and $D_2$ at the boundary $z=0$ are related by the matching conditions of the fields $B$ and $\phi$ at a matching point $z_m$.

\begin{figure}[H]
\centering
\includegraphics[scale=0.4]{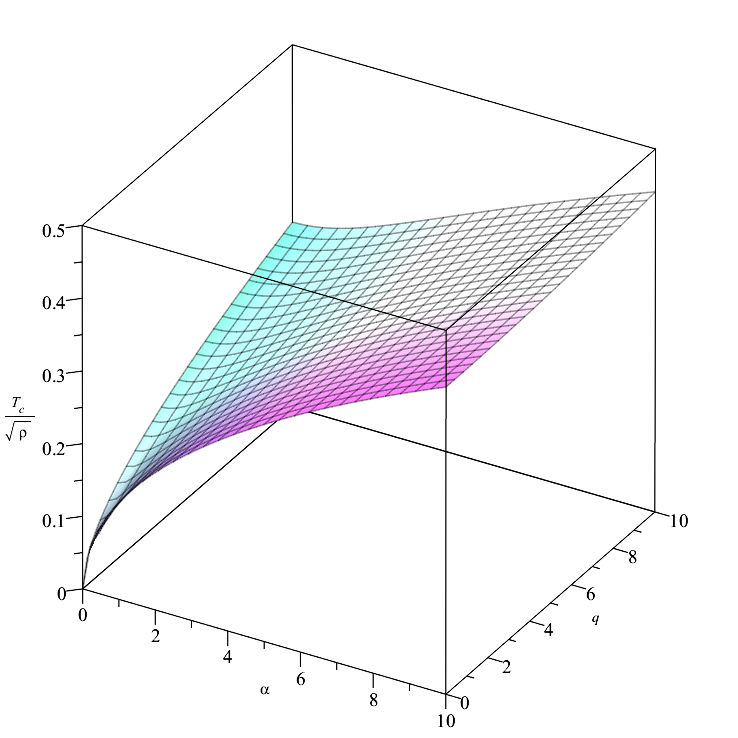}
\caption{$T_{c}/\sqrt{\rho}$ varies with the couplings $\alpha$, $q$.} \label{critical temperature}%
\end{figure}

To match the series expansions of the fields $B(z)$ and $\phi(z)$ at the horizon in Eq. (\ref{fH}) and that at the boundary in Eq. (\ref{fB}) smoothly at a matching point $z_{m}$, we require the following four constraint equations,

\begin{subequations}
\begin{align}\label{meq-mfa}
  \mu - \frac{\rho}{r_H}z_{m}  &  = a\left( 1-z_{B} \right) + B_{2}\left( 1-z_{m} \right)^{2},\\
  -\frac{\rho}{r_H}  &  = - a - 2B_{2}\left( 1-z_{B} \right),\\
  \frac{D_2}{r_H^2}z_{m}^{2}  &  = b + \phi_{1}\left( 1-z_{m} \right) + \phi_{2}\left( 1-z_{m} \right)^{2},\\
  \frac{2D_2}{r_H^2}z_{m}  &  = -\phi_{1} - 2\phi_{2}\left( 1-z_{m} \right).
  \label{meq-mfd}
\end{align}
\end{subequations}
The above constraint equations can be solved as,
\begin{subequations}
\begin{align}
  \mu  &  = a + B_{2}\left( 1-z_{m} \right) \left( 1+z_{m} \right),\\
  \rho  &  = \left[ a + 2B_{2}\left( 1-z_{m} \right)  \right] r_{H}, \label{eqab-non-1}\\
  D_{2}  &  = \frac{b }{ 12z_{m} } \left[ 12r_H^2-(\alpha^2 a^2+4r_H^2)(1-z_m) \right],\label{D+}\\
  0  &  = 2b + \phi_{1}\left( 2-z_{m} \right) + 2\phi_{2}\left( 1-z_{m} \right),
\label{eqab-non-2}%
\end{align}
\end{subequations}
To be concrete, we take $z_{m}=3/4$ in the following calculation. The two parameters $a$ and $b$ can be analytically solved from Eqs. (\ref{eqab-non-1}) and (\ref{eqab-non-2}) once the charge density $\rho$ is given,
\begin{align}
  {b}^{2}=&{\frac{-24{r_H}(a{r_H}-\rho)}{2a(8{\alpha}^{2}+{q}^{2})r^2_H-12{\alpha}^{2}{r_H}\rho+{\alpha}^{4}{a}^{3}}},
  \label{b2}
\end{align}
with $a$ satisfies a quartic equation,
\begin{align}
  {\alpha}^{4}{a}^{4}-4{r_H^2(-2{\alpha}^{2}+{q}^{2}){a}^{2}-48{r_H}\rho{\alpha}^{2}a+304r_H^4}=&0.
  \label{a4}
\end{align}
The condensate $D_{2}$ is the order parameter of the superconducting phase transition. Above a critical temperature $T_{c}$, the order parameter $D_{2}\propto b$ is zero, which from Eq. (\ref{b2}) implies $ar_{H}-\rho=0$ at the critical temperature $T_{c}$.  From Eq. (\ref{a4}), we obtain the critical temperature by taking  $a=\rho/r_H$,
\begin{equation}
  T_{c} = \frac{3}{4\pi} \left( \frac{ 10\alpha^2 + q^2 + \sqrt{24\alpha^4 + 20\alpha^2q^2 + q^4} }{ 152 } \right)^{1/4} \rho^{1/2}.
  \label{Tc}
\end{equation}
For $\alpha=q=0$, $T_c$ vanishes, there is no phase transition at this special point. For $\alpha\gg q$, the critical temperature $T_c\propto \alpha^{1/2}$; While for $\alpha\ll q$, the critical temperature $T_c\propto q^{1/2}$. The critical temperature $T_c$ is proportional to $\sqrt{\rho}$ and increases with $\alpha$ and $q$ monotonously. $T_{c}/\sqrt{\rho}$ vs. the parameters $\left(\alpha,q\right) $ is plotted in Fig. \ref{critical temperature}. The behavior of the analytic expression of the critical temperature in Eq. (\ref{Tc}) is consistent with the numeric result in \cite{1006.2726}.

Near the critical temperature $T_c$, the condensate behaves as $D_{2}=AT_c^2(1-T/T_c)^{1/2}$, which indicates that the phase transition at $T_c$ is a second order phase transition. The coefficient $A$ depends on the parameters $\alpha$, $q$ and $\rho$, see Eq. (\ref{D_Tc_n}) in the Appendix.

\begin{figure}[H]
\centering
\includegraphics[scale=0.4]{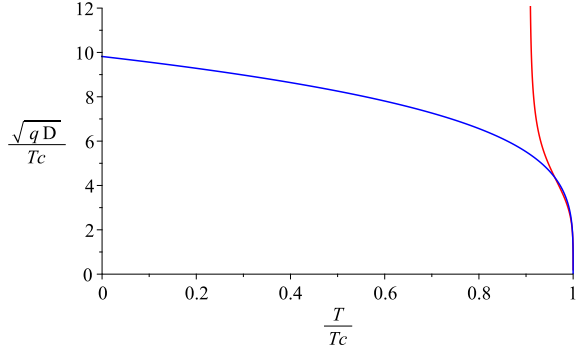}
\caption{The behavior of the condensate $D_{2}$ with the couplings $\alpha=1$, $q=1$. The red line is the condensate directly from Eq. (\ref{D+}), while the blue line is the approximate condensate near the critical temperature $T_c$ .} \label{D+T}%
\end{figure}

The condensate $D_{2}$ vs. $T/T_c$ for $\alpha=q=1$ can be calculated from Eq. (\ref{D+}) and is plotted as the solid line in Fig. \ref{D+T}. We see that the low temperature behavior of condensate in the non-backreaction calculation (red line) is not consistent with the numeric calculation in \cite{1006.2726}. The blue line in Fig. \ref{D+T} is the condensate near the critical temperature and is exploited to the low temperature region.
The Fig. \ref{D+T} shows the drawback of the non-backreaction approximation that is also seen in \cite{1211.0904}.
This drawback seeks an improvement on the ordinary matching method.

The inconsistency of the condensate around the low temperature is due to the non-backreaction approximation we took in this section. In the next section, we will study the full backreaction case.

\setcounter{equation}{0} \renewcommand{\theequation}{\arabic{section}.\arabic{equation}}

\section{Full-Backreaction}

In this section, we consider the case of full-backreaction by including the backreacted effects of the fields $B$ and $\phi$.
It is necessary to solve the fields $\chi$ and $g$ in the metric (\ref{metric}) as well as the fields $B$ and $\phi$ together from the full equations of motion (\ref{eom-full}).

Similarly to the non-backreaction case, there are three scaling symmetries in the equations of motion (\ref{eom-full}) in the full-backreaction case.
The scaling dimensions of these variables for the scaling symmetries are listed in Table \ref{scaling-full}. 

\begin{table}[H]
\centering
\scalebox{1}{
\begin{tabular}[c]{ |c|c|c|c|c|c|c|c|c|c| } \hline
  Sym. & $n_r$ & $n_t$  & $n_{\vec{x}}$ & $n_{r_{H}}$ & $n_{e^{\chi}}$ & $n_{g}$ & $n_{B}$& $n_{\ell}$& $n_{q}$\\ \hline
  I   &  1  &  1  &  0  &  0  &  2  & 0 &  0  & 0 &  0 \\ \hline
  II  &  1  & -1  &  -1 &  1  &  0  & 2 &  1  & 0  &  0 \\ \hline
  III &  1 &  1  &  0  &  1  &  0  & 0 &  0  & 1  & -1 \\ \hline
\end{tabular}}
\caption{Scaling symmetries in the equations of motion (\ref{eom-full}) with full-backreaction.
} \label{scaling-full}%
\end{table}
As in the non-backreaction case, we can use the three scaling symmetries to set the parameters $\ell=1$, $\rho=$ constant and $\chi^{(0)}=0$ in Eq. (\ref{chi}), which are the necessary boundary conditions to ensure the asymptotic AdS at the boundary.

\subsection{Generalized Matching Solutions}

Similarly as what we have done in the non-backreaction case, we will solve the fields $(\chi, g, B, \phi)$ in the full-backreaction case by using the matching method.

Up to the order next to the initial values, the series expansions of the  fields $(\chi, g, B, \phi)$ near horizon in (\ref{fHfull}) becomes,
\begin{subequations}
\label{fHfull_}
\begin{align}
  \chi\left(z\right)  &  = \chi_{0} + \chi_{1}\left(1-z\right),\\
  g\left(z\right)  &  = \frac{r_H^2}{z^2} \left[ 1 - z^3 + g_1\left(1-z\right) \right],\\
  B\left(z\right)  &  = a\left(1-z\right) + B_2\left(1-z\right)^2,\\
  \phi\left(z\right)  &  = b + \phi_1\left(1-z\right) + \phi_2\left(1-z\right)^2,
\end{align}
\end{subequations}
where we have renamed $B_1=a$ and $\phi_0=b$ in (\ref{fHfull}). In the above expansion, $\chi_0$, $a$ and $b$ need to be imposed as the boundary values at the horizon $z= 1$. Plug the series expansions (\ref{fHfull_}) into the equations of motion (\ref{eom-full}), we can solve for the coefficients $(\chi_1, g_1, B_2, \phi_2)$ in Eqs. (\ref{fHfull_}) in terms of the boundary values $(\chi_0, a, b)$.
\begin{subequations}
\begin{align}
g_1&=\frac{1}{8 r_H^2} \Big( 4b^2 r_H^2-a^2(\alpha^2 b^2+2) \Big),\\
\phi_1&=-\frac{(a^2\alpha^2+4r_H^2)b}{2(g_1+3)r_H^2},\\
\chi_1&=\frac{-\phi_1^2(g_1+3)^2 r_H^2-a^2}{2(g_1+3)^2 r_H^2},\\
B_2 &= \frac{a \Big[ \left( \frac{1}{4}\chi_1(g_1+3)\alpha^2+q^2 \right) b^2-\phi_1\alpha^2(g_1+3)b+\frac{1}{2}(g_1+3)\chi_1 \Big]}{(\alpha^2 b^2+2)(g_1+3)},\\
\phi_2&=\frac{1}{8(g_1+3)^2 r_H^2} \bigg\{ \bigg[ \Big( \big( (\chi_1+4)b-\phi_1 \big) g_1+(3\chi_1+15)b-3\phi_1 \Big)\alpha^2-2b q^2 \bigg] a^2 - 4b B_2\alpha^2(g_1+3)a\notag\\
&+12r_H^2 \left[\frac{1}{12} g_1^2\chi_1\phi_1+\frac{1}{2}\left(\chi_1+\frac{7}{3}\right)\phi_1 g_1+b+\phi_1\left(\frac{3}{4}\chi_1+\frac{7}{2}\right) \right] \bigg\}.
\end{align}
\end{subequations}
The detailed forms are listed in (\ref{fBhc}).

At the boundary $z=0$, the asymptotic behavior of the fields $(\chi, g, B, \phi)$ was listed in Eq. (\ref{fBfull}), where $M$, $\mu$ and $D_2 $ will be imposed as the boundary values at the boundary $z= 0$. 

The three boundary values $(\chi_0, a, b)$ at the horizon $z=1$ and the three boundary values $(M, \mu, D_2)$ at the boundary $z=0$ are related by six matching conditions of the fields $(\chi, g, B, \phi)$ at a matching point $z_{m}$.

We match the asymptotic forms of the fields at the horizon in Eq. (\ref{fBfull}) and that at the boundary in Eq. (\ref{fHfull_}) at a matching point $z_{m}$.
Besides the four matching equations (\ref{meq-mfa} -\ref{meq-mfd}) for the matter fields as we listed in the non-backreaction case, additional matching constraints are needed for the gravitational fields.
The equations of motion for the gravitational fields $\chi(z)$ and $g(z)$ are first order differential equations as shown in Eq. (\ref{g1}) and Eq.(\ref{g2}), so that we can only assign two boundary conditions.
Therefore, only two more continuous matching conditions are applied on the gravitational fields
\begin{subequations}
\begin{align}\label{meq-ga}
  \frac{1}{2} \left( \frac{D_2}{r^2_H}z_{m}^2 \right)^2  &  = \chi_{0} + \chi_{1} \left(1-z_{m}\right),\\
  \frac{1-2Mz_{m}^3}{z_{m}^2}r^2_H  &  = \frac{1-z_{m}^3}{z_{m}^2}r^2_H + g_{1}\frac{\left(1-z_{m}\right)}{z_{m}^2}r^2_H,\label{meq-gb}
\end{align}
\end{subequations}
and we have to give up the smooth conditions.

The total six matching equations  (\ref{meq-mfa} -\ref{meq-mfd}) and  (\ref{meq-ga} -\ref{meq-gb}) can be solved as
\begin{subequations}
\begin{align}
  \chi_{0}  &  = -\chi_{1} \left(1-z_{m}\right) + \frac{1}{2} \left( \frac{D_2}{r^2_H} \right)^2z_{m}^4,\\
  M  &  = \frac{1}{2} - \frac{ g_{1}\left(1-z_{m}\right) }{ 2z_{m}^3 },\\
  \mu  &  = a + B_{2} \left(1-z_{m}\right) \left(1+z_{m}\right),\\
  \rho  &  = \Big[ a + 2B_{2}\left(1-z_{m}\right) \Big]r_H,\label{eqrho}\\
  D_2  &  = \frac{ 2b + \phi_{1}\left(1-z_{m}\right) }{ 2z_{m} }r^2_H,\\
  0  &  = 2b + \phi_{1} \left(2-z_{m}\right) + 2\phi_{2} \left(1-z_{m}\right).
  \label{eqab}%
\end{align}
\end{subequations}
The same as in the non-backreaction case, we take $z_{m}=3/4$ in the following.
Eq.(\ref{eqab}) can be expressed as a cubic equation of $a^{2}$, see Eq. (\ref{y3}), which can be solved analytically in term of $b^{2}$.

\begin{figure}[H]
\centering
{\includegraphics[scale=0.4]{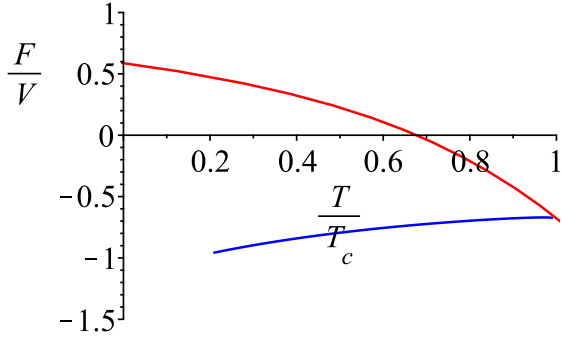}}
\caption{The free energies $\frac{F_{Hairy}}{V}$(blue) and $\frac{F_{RN}}{V}$(red) with $\alpha=5$, $q=1$, and $\rho=$ constant for $T\le T_{c}$.}
\label{free energy}%
\end{figure}

We check the stability of the analytic hairy black hole solution by comparing its free energies with the free energies of RN black holes are obtained in Eqs. (\ref{RN}),%
\begin{subequations}
\begin{align}
  F_{Hairy}  &  = V\left(-2M+\mu\rho\right),\\
  F_{RN}  &  = \frac{V}{r_H} \left(-r^4_H + \frac{3\rho^2}{4}\right).
\end{align}
\end{subequations}
For the given couplings $\alpha$ and $q$, as well as a charge density $\rho$, the free energies $F_{Hairy}$ and $F_{RN}$ vs. $T/T_c$ are plotted in Fig. \ref{free energy}.
When $T<T_{c}$, the free energy of the hairy black hole is always lower than that of the RN black hole and vice versa. Thus for  $T>T_{c}$,  the system is in the RN black hole phase with the scalar field $\phi=0$; while for $T<T_{c}$, the system transits to the hairy black hole phase with $\phi\ne0$ that indicates the condensation.

\subsection{Condensate}

\begin{figure}[H]
\centering
\subfloat[]{\includegraphics[scale=0.4]{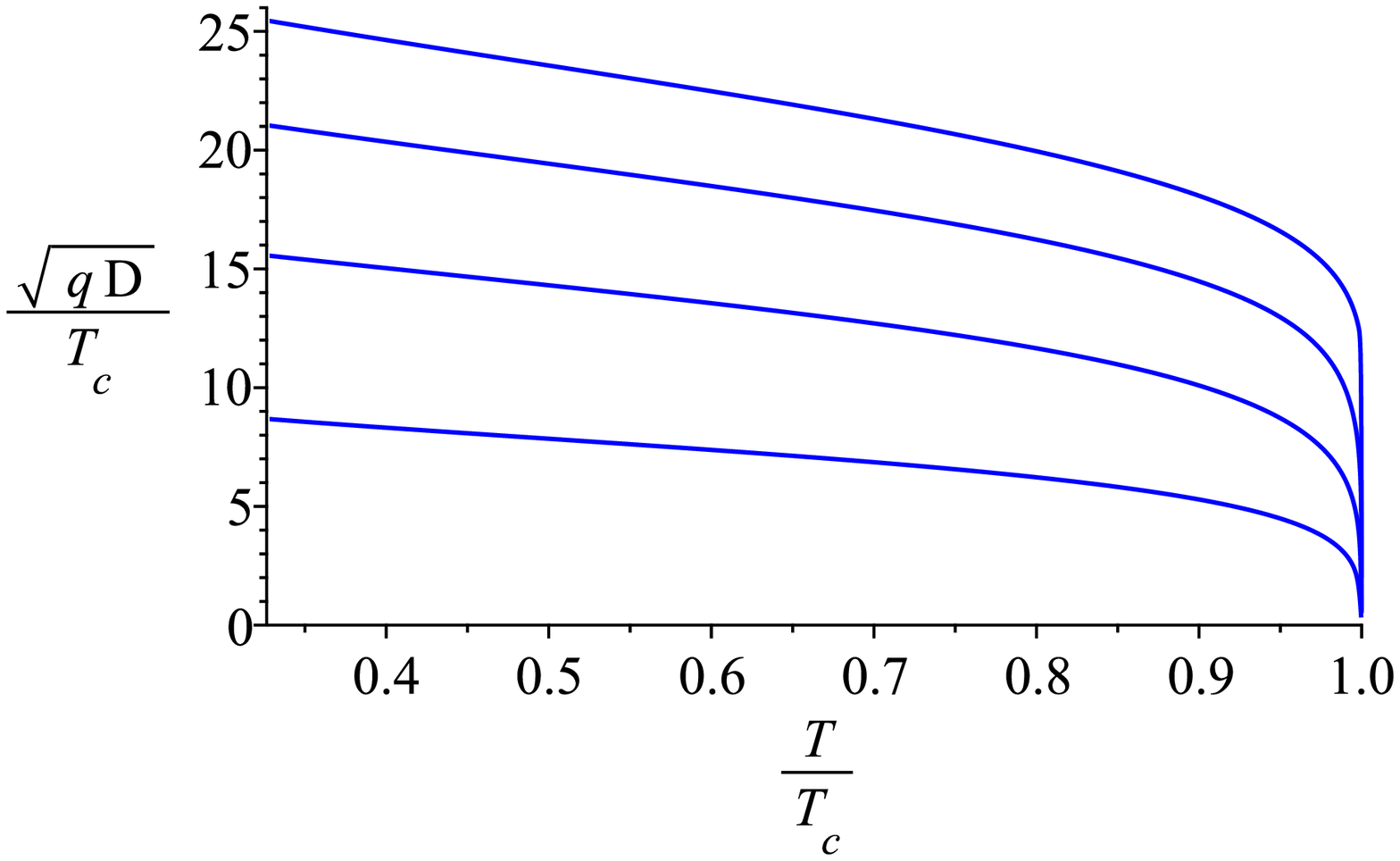}}
\subfloat[]{\includegraphics[scale=0.4]{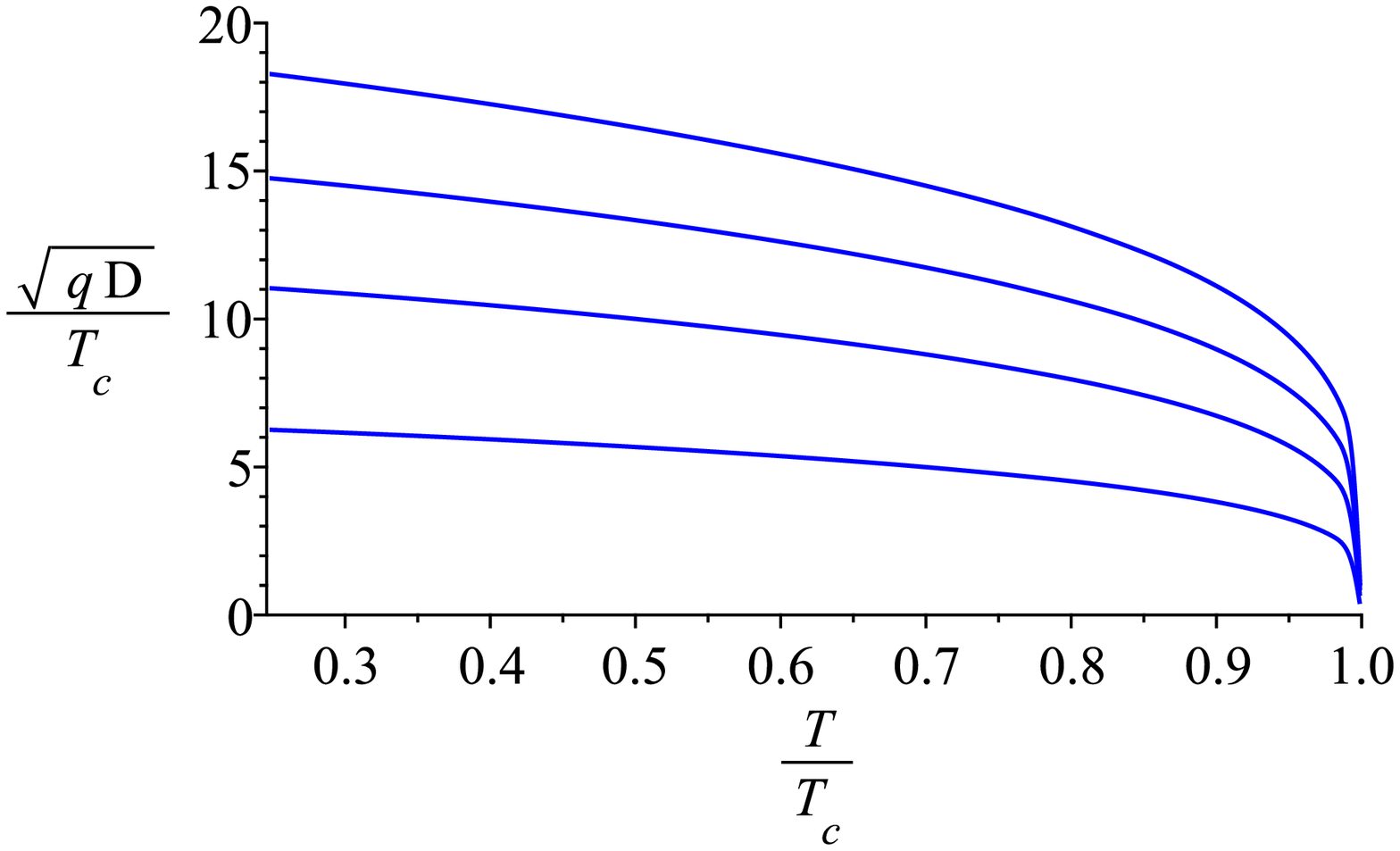}}
\caption{The condensate to temperature diagram with the coupling $q=1,3,5,7$ from bottom to top and (a) $\alpha=1$, (b) $\alpha=5$.}%
\label{condensate q}%
\end{figure}

\begin{figure}[H]
\centering
\subfloat[]{\includegraphics[scale=0.4]{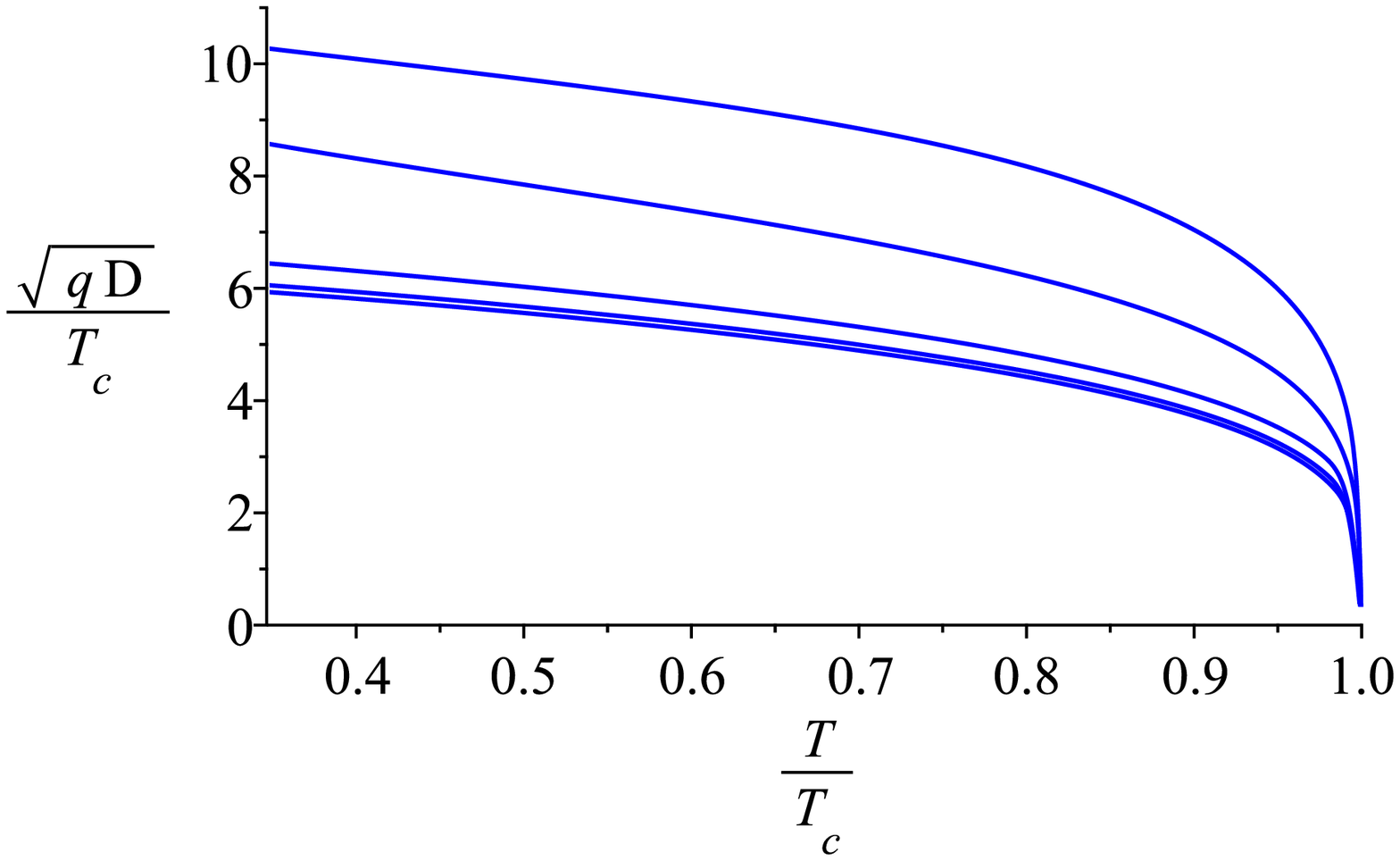}}
\subfloat[]{\includegraphics[scale=0.4]{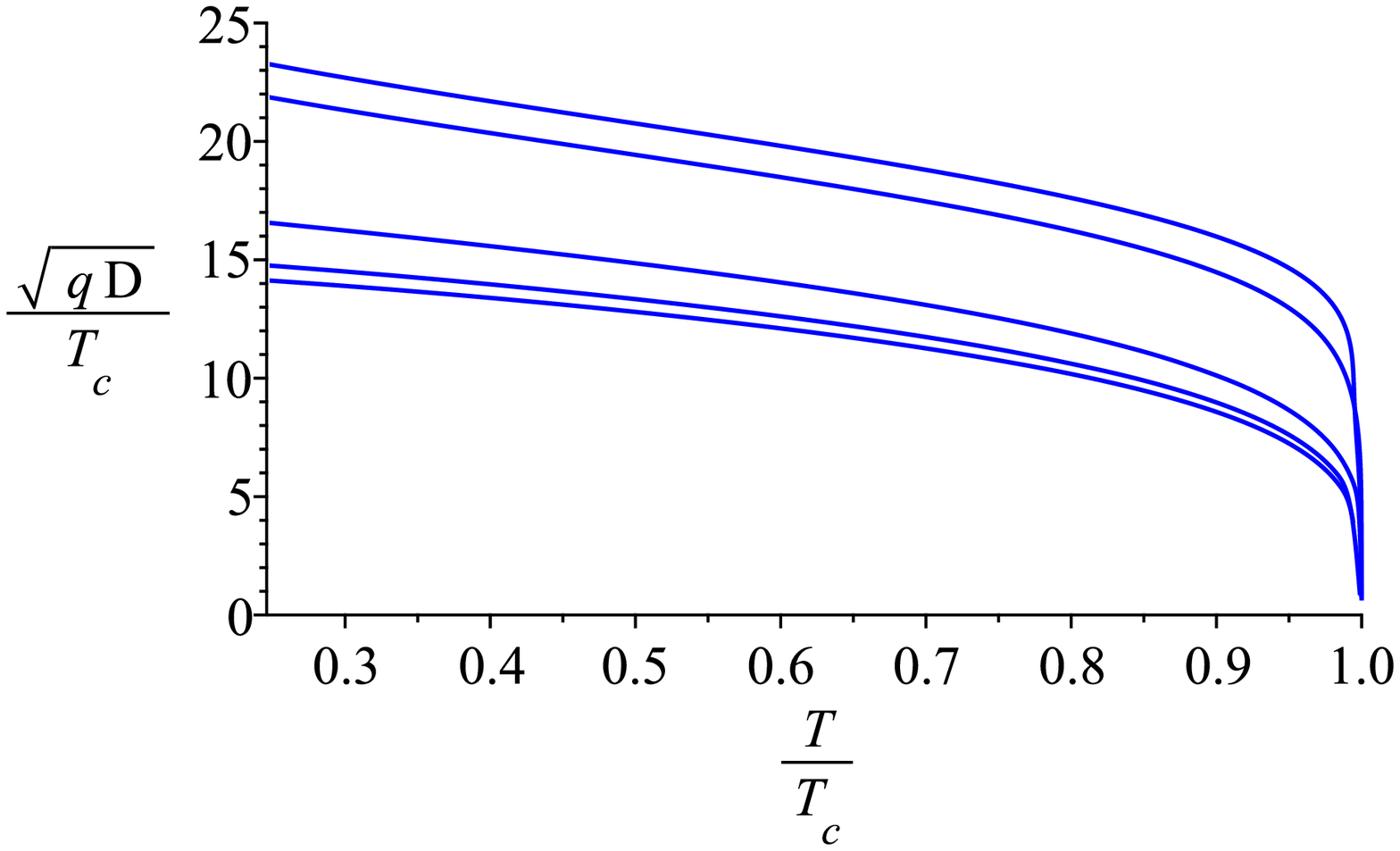}}
\caption{The condensate to temperature diagram with the coupling $\alpha=10^{-4},1,3,5,7$ from top to bottom and (a) $q=1$, (b) $q=5$.}%
\label{condensate alpha}%
\end{figure}

In this section, we investigate the condensate $D_2$ , the order parameter of the superconducting phase transition, in more details.
The behavior of the order parameter $D_2\propto b$ around the critical temperature is the same as that in the non-backreaction case, $D_2\propto \left(1-T/T_c\right)^{1/2}$. However, the fixed-$\rho$ condition has been modified at the critical temperature ($B_2|_{T=T_c} \neq 0$).
The condensate $D_{2}$ vs the ratio of the temperature $T/T_c$ for different parameters $\alpha$ and $q$ are plotted in Fig. \ref{condensate q} and \ref{condensate alpha}.
Form Fig. \ref{condensate q} and Fig. {condensate alpha}, we see that, for a fixed $\alpha$, the condensate $D_2$ increases as the parameter $q$ increases, while for a fixed $q$, the condensate $D_2$ decreases as the parameter $\alpha$ increases. The behavior of the condensate $D_2$ depending on the parameters $\alpha$ and $q$ is consistent with the numeric result in \cite{1006.2726}.

\begin{figure}[H]
\centering
\includegraphics[scale=0.4]{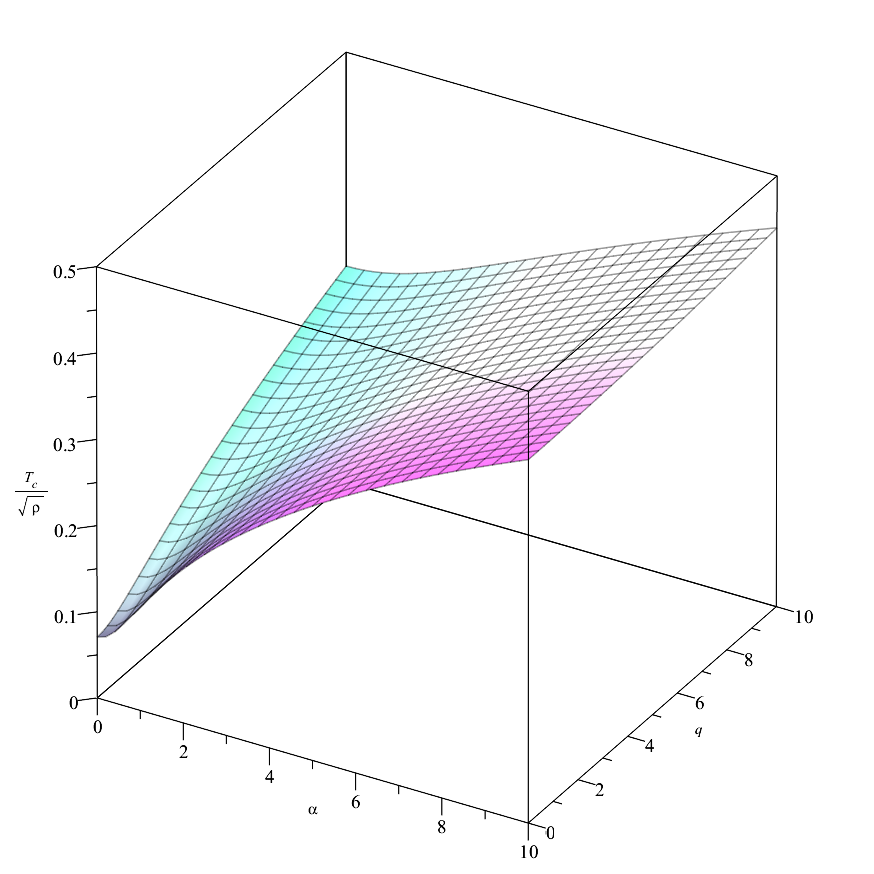}
\caption{The critical temperature $T_{c}$ vary with the couplings $\alpha$, $q$.}
\label{critical temperature full}%
\end{figure}

As we did in the non-backreaction case, the critical temperature in the full-backreaction case can be obtained analytically, but the expression is much more complicated.
The behavior of the critical temperature $T_c$ in the full-backreaction case is also similar to that in the non-backreaction case and is plotted in Fig. \ref{critical temperature full}.

\subsection{Conductivity}

In this section, we compute the conductivity of this system by adding an external source $A_{x}$ perturbatively.
The metric is thus modified by a perturbative non-diagonal component $g_{tx}$. Defining $A_{x}=\tilde{A}_{x}\left(r\right)e^{-i\omega t}$ and $g_{tx}=\tilde{g}_{tx}\left(r\right)e^{-i\omega t}$, the equations of motion for the perturbative field $\tilde{A_{x}}\left(r\right)$ and $\tilde{g}_{tx}\left(r\right)$ are
\begin{subequations}
\begin{align}
  \tilde{A}_{x}^{\prime\prime} + \left( \frac{g^{\prime}}{g} - \frac{\chi^{\prime}}{2} + \frac{f^{\prime}}{f} \right) \tilde{A}_{x}^{\prime} + \left( \frac{\omega^{2}}{g^{2}}e^{\chi} - \frac{J}{gf} \right) \tilde{A}_{x} + \frac{A_{t}^{\prime}}{g}e^{\chi} \left( \tilde{g}_{tx}^{\prime} - \frac{2}{r}\tilde{g}_{tx} \right)  &  = 0,\\
  \tilde{g}_{tx}^{\prime} - \frac{2}{r}\tilde{g}_{tx} + fA_{t}^{\prime}\tilde{A}_{x}  &  = 0,
\end{align}
\end{subequations}
which lead to a homogeneous linear differential equation for $\tilde{A}_{x}$,
\begin{equation}
  \tilde{A}_{x}^{\prime\prime} + \left( \frac{g^{\prime}}{g} - \frac{\chi^{\prime}}{2} + \frac{f^{\prime}}{f} \right) \tilde{A}_{x}^{\prime} + \left[ \left( \frac{\omega^{2}}{g^{2}} - \frac{fA_{t}^{\prime2}}{g} \right) e^{\chi} - \frac{J}{gf} \right] \tilde{A}_{x} = 0.
\end{equation}
Making a coordinate transformation $z=r_{H}/r$ and defining $e^{\chi}\tilde{A}_{x}^{2}=C^{2}$, the equation  for $\tilde{A}_{x}$ becomes
\begin{align}
  C^{\prime\prime} + \left( \frac{g^{\prime}}{g} - \frac{3}{2}\chi^{\prime} + \frac{f^{\prime}}{f} + {\frac{2}{z}} \right) C^{\prime}  &  \notag\\
    + \left[ -\frac{1}{2}\chi^{\prime\prime} + \frac{1}{2}\chi^{\prime 2} - \frac{1}{2}\chi^{\prime} \left( \frac{g^{\prime}}{g} + {\frac{f^{\prime}}{f}} + {\frac{2}{z}} \right) + \left( \frac{r^2_H\omega^2e^\chi}{z^4g^2} - \frac{f}{g} \left( B^{\prime} - \frac{1}{2}\chi^{\prime}B \right)^2 - \frac{r^2_HJ}{z^4gf} \right) \right] C  &  = 0.
\label{eom-c}
\end{align}
To realize the structure of the Eq. (\ref{eom-c}), we make a further coordinate transformation,
\begin{equation}
    du=-\frac{r_H e^{\frac{\chi}{2}}}{g z^2}dz,
    \label{coordinate trans}
\end{equation}
which transforms the horizon at $z=1$ to $u=-\infty$ and the boundary at $z=0$ to $u=0$.

We integrate the coordinate transformation (\ref{coordinate trans}) term by term to get
\begin{equation}
    u=-\int \left(\frac{p_0}{1-z}+p_1+...\right)dz=p_0 \ln (1-z)+p_1(1-z)+\cdots,
\end{equation}
where $p_i$'s are the expansion coefficients near the horizon at $z=1$. 

By defining a new field $\Psi=\sqrt{f}e^{-\chi/2}C$, the Eq.(\ref{eom-c}) can be brought to the form of the Schrodinger equation,
\begin{equation}
   \frac{d^2 \Psi}{du^2}+[\omega^2-V(u)]\Psi=0,
   \label{Schrodinger}
\end{equation}
where the potential is
\begin{equation}
   V(u)=g \left[ f \left(\frac{d(e^{-\chi/2} B)}{du} \right)^2+\frac{J}{f}e^{-\chi} \right]+\frac{1}{\sqrt{f}}\frac{d^2 \sqrt{f}}{du^2},
   \label{potential}
\end{equation}
with $V(0)=V(-\infty)=0$.

The Schrodinger equation (\ref{Schrodinger}) with the potential (\ref{potential}) is a standard one-dimensional scattering problem, and the wave function $\Psi(u)$ near the boundary at $u\sim0$ behaves as
\begin{equation}
   \Psi(u)=e^{-i\omega u}+R e^{i\omega u},
\end{equation}
where $R$ is the reflection coefficient. The conductivity can thus be written as
\begin{equation}
   \sigma(\omega)=\frac{1-R}{1+R}.
\end{equation}
At the horizon at $u=-\infty$, the in-falling wave boundary condition admits a non-reflection wave function ,
\begin{equation}
   \Psi(u)=Te^{-i\omega u}.\label{infall-u}
\end{equation}
where $T$ is the transmission coefficient.

In the $z$ coordinate, the in-falling wave function becomes
\begin{equation}
   \Psi=Te^{-i\omega [p_0 \ln (1-z)+p_1(1-z)+\cdots]}.
   \label{infall-z}
\end{equation}
Therefore, near the horizon, the field $C\left(  z\right)  $ can be expanded as,
\begin{align}
   C(z)&= c_0e^{ -i\omega [p_{0}\ln{(1-z)}+p_{1}(1-z)+\cdots] } \cdot  \big[ 1 + C_{1}\left(1-z\right)+\cdots \big] \notag \\
   &\simeq c_0e^{ -i\omega [p_{0}\ln{(1-z)}+P_{1}(1-z)] } \cdot \big[ 1 + C_{1}\left(1-z\right) \big],
   \label{c}
\end{align}
where in the second line we assumed a truncated form of the field $C(z)$ with $P_1$ and $C_1$ being the effective coefficients by truncating all the higher-order terms at the horizon. Therefore, $P_1$ and $C_1$ contain the effects far from the horizon and will be determined by Eq. (\ref{c}) and the boundary condition at $z=0$. The truncating solution is an approximation of the true solution, but we will see that many important properties are preserved in this approximation.

At the boundary, the asymptotic form of $C\left(  z\right)  $ is
\begin{equation}
  C\left(z\right) = C^{ \left(0\right) } + \frac{ C^{\left(1\right)} }{ r_H }z + \cdots.
\end{equation}
Expanding the field $C(z)$ in Eq. (\ref{c}) at the boundary $z=0$, the conductivity can be calculated as follows \cite{0810.1563},
\begin{align}
  \sigma\left(\omega\right) =& \frac{1}{i\omega} \frac{C^{\left(1\right)}}{C^{\left(0\right)}} =r_H(p_0+P_1)- \frac{r_{H}}{i\omega} \left( \frac{C_{1}}{1+C_{1}} \right).
  \label{conductivity}
\end{align}
In the high frequency limit $\omega\rightarrow \infty$, the conductivity should approach to one due to the asymptotic AdS geometry, the above expression gives
\begin{align}
   P_{1} =& -p_{0} + \frac{1}{r_{H}}.
   \label{P1}
\end{align}
The coefficients $p_{0}$ and $C_{1}$ can be now calculated by plugging the the field $C(z)$ in Eq. (\ref{c}) into the equation of motion (\ref{eom-c}). The exact expressions of $p_{0}$ and $C_{1}$ are listed in the Appendix as Eqs. (\ref{p0}) and (\ref{c1})..

The DC conductivity can be obtained from the low frequency expansion of the conductivity (\ref{conductivity}),
\begin{align}
  \sigma \left(\omega\right) &= \Re(\sigma)+i\Im(\sigma), \\
\Re(\sigma)&=\sigma_{0}+ O\left(\omega^2\right),\\
\Im(\sigma)&=\sigma_{-1}\omega^{-1}+ O\left(\omega\right).
\end{align}
where both $\sigma_0$ and $\sigma_{-1}$ are real and they are given in Eq. (\ref{sigDC}) in the Appendix. At $\omega=0$, the imaginary part of the conductivity $\Im(\sigma)$ is proportional to $\omega^{-1}$. By Kramers-Kronig relations, this implies that the real part of the conductivity $\Re(\sigma)$ behaves as a delta function at $\omega=0$, i.e. DC superconductivity.

In the high frequency, the conductivity can be expanded as,
\begin{equation}
  \sigma\left(\omega\right) = 1 + \frac{ir_H}{\omega} + O\left(\omega^{-2}\right),
\end{equation}
which gives $\sigma(\omega)\rightarrow 1$ as $\omega\rightarrow \infty$ that has been fixed  by the choice of $P_1$ in Eq. (\ref{P1}).

\begin{figure}[H]
\centering
\begin{table}[H]
\centering
\begin{tabular}[c]{ |c|c|c| } \hline
\centering
\includegraphics[width=0.3\textwidth]{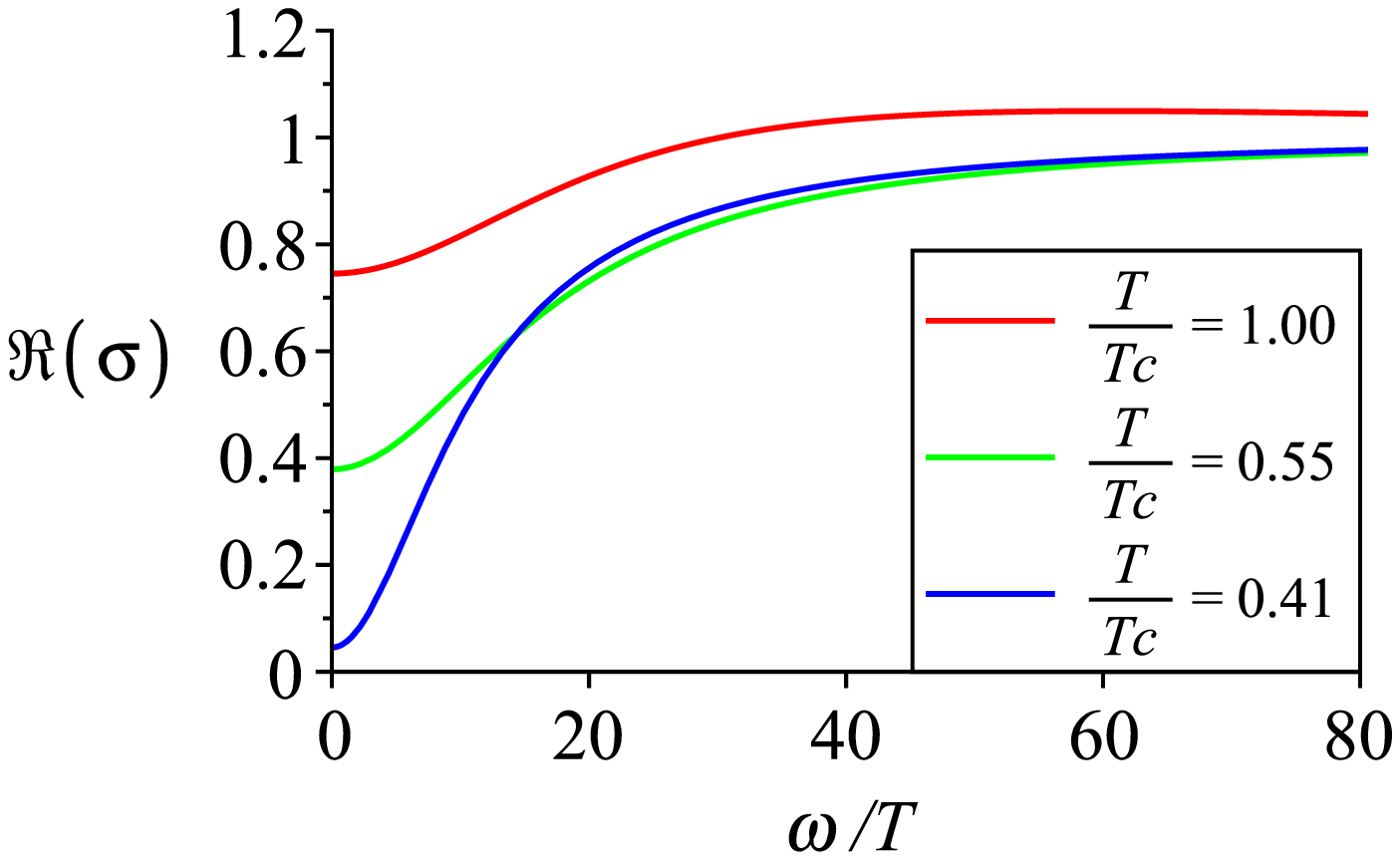}&
\includegraphics[width=0.3\textwidth]{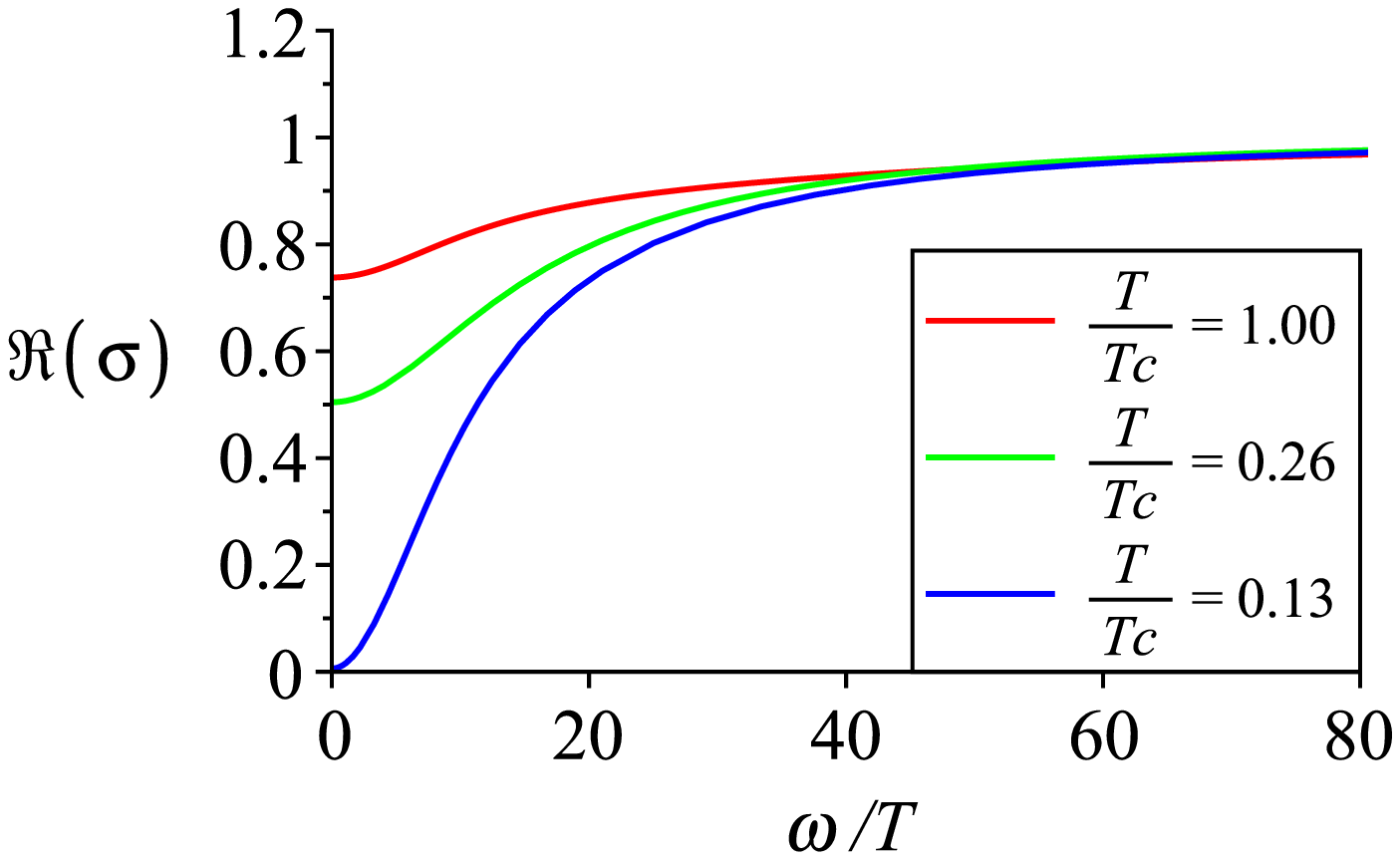}&
\includegraphics[width=0.3\textwidth]{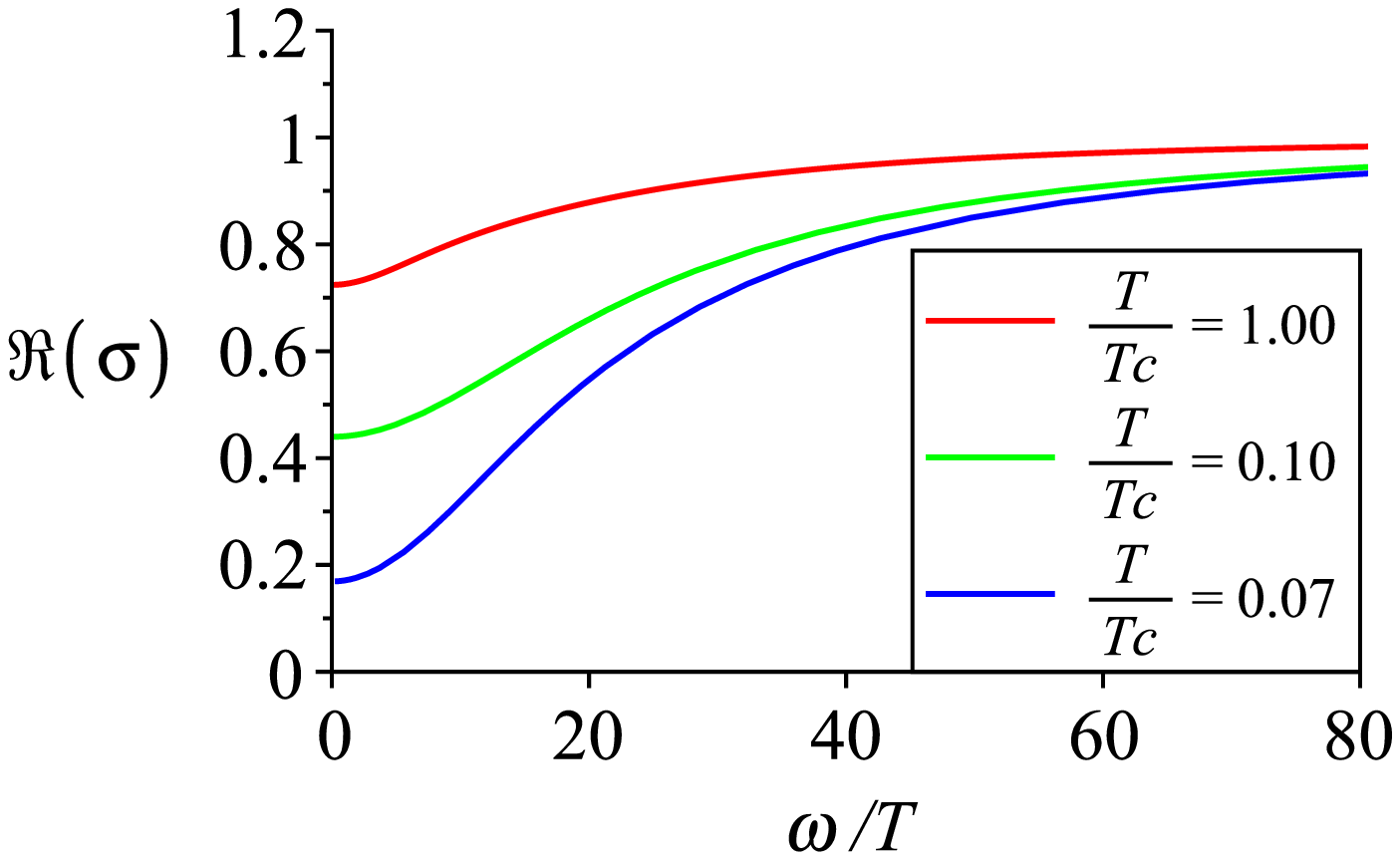}
\\ \hline
\includegraphics[width=0.3\textwidth]{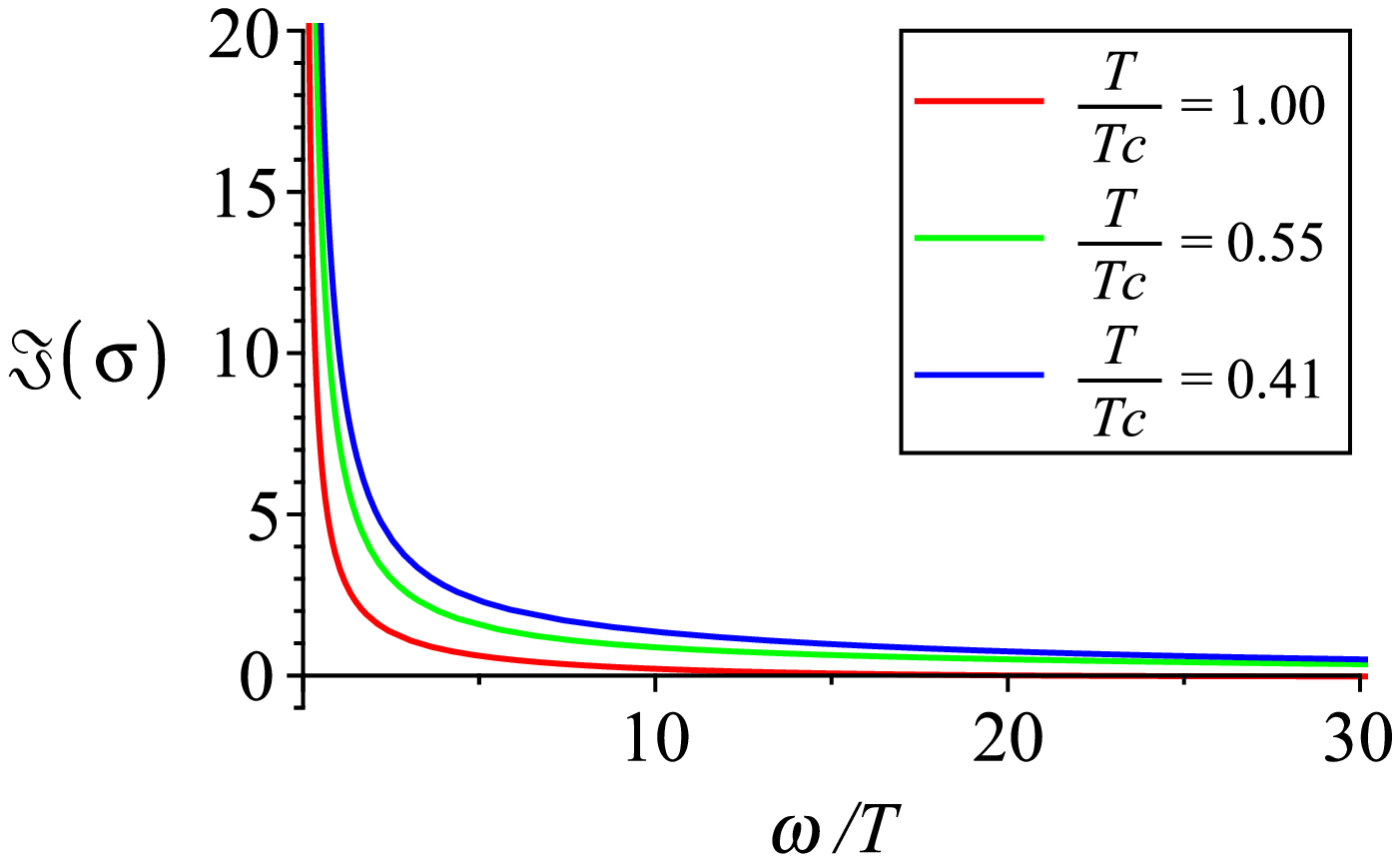}&
\includegraphics[width=0.3\textwidth]{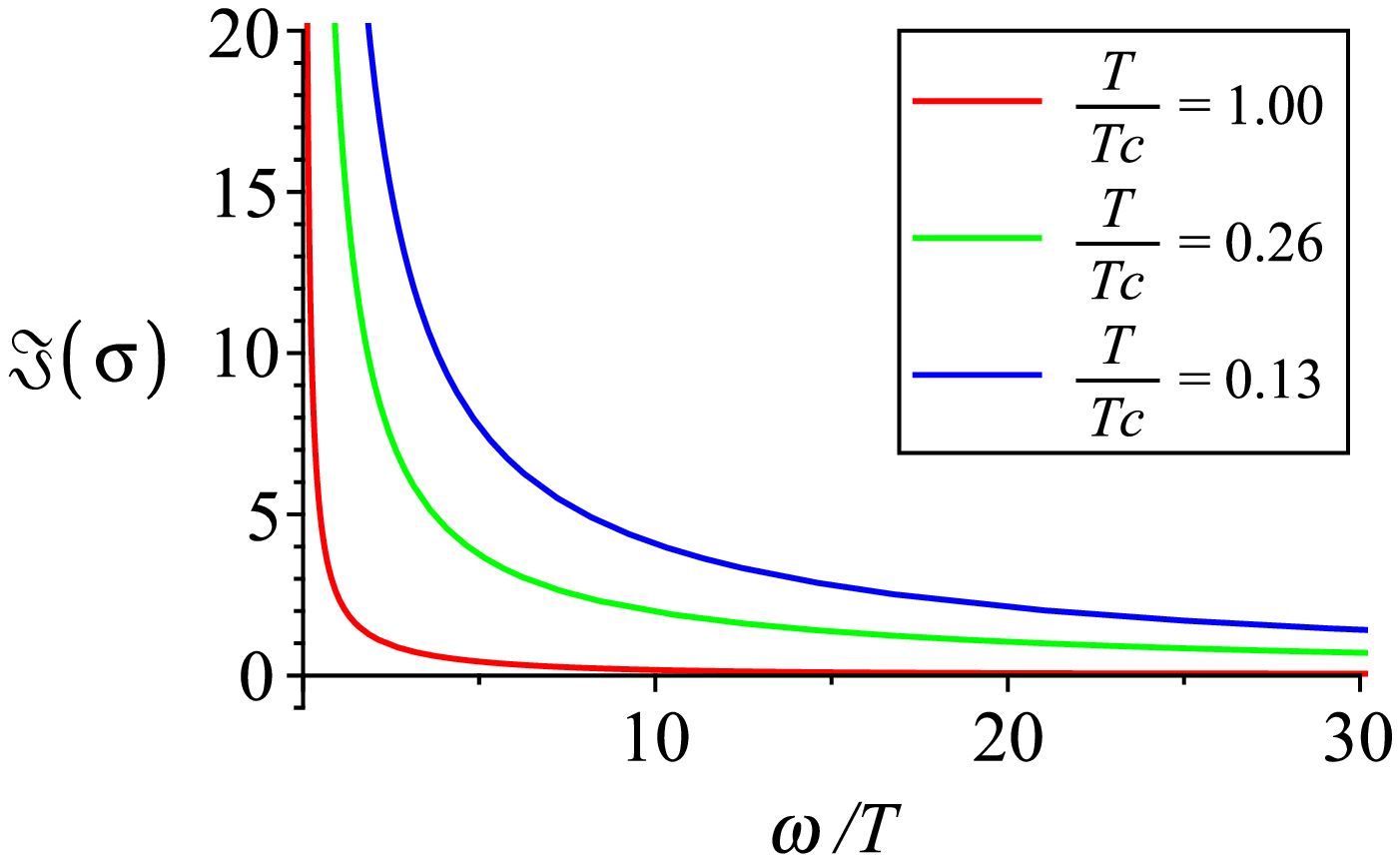}&
\includegraphics[width=0.3\textwidth]{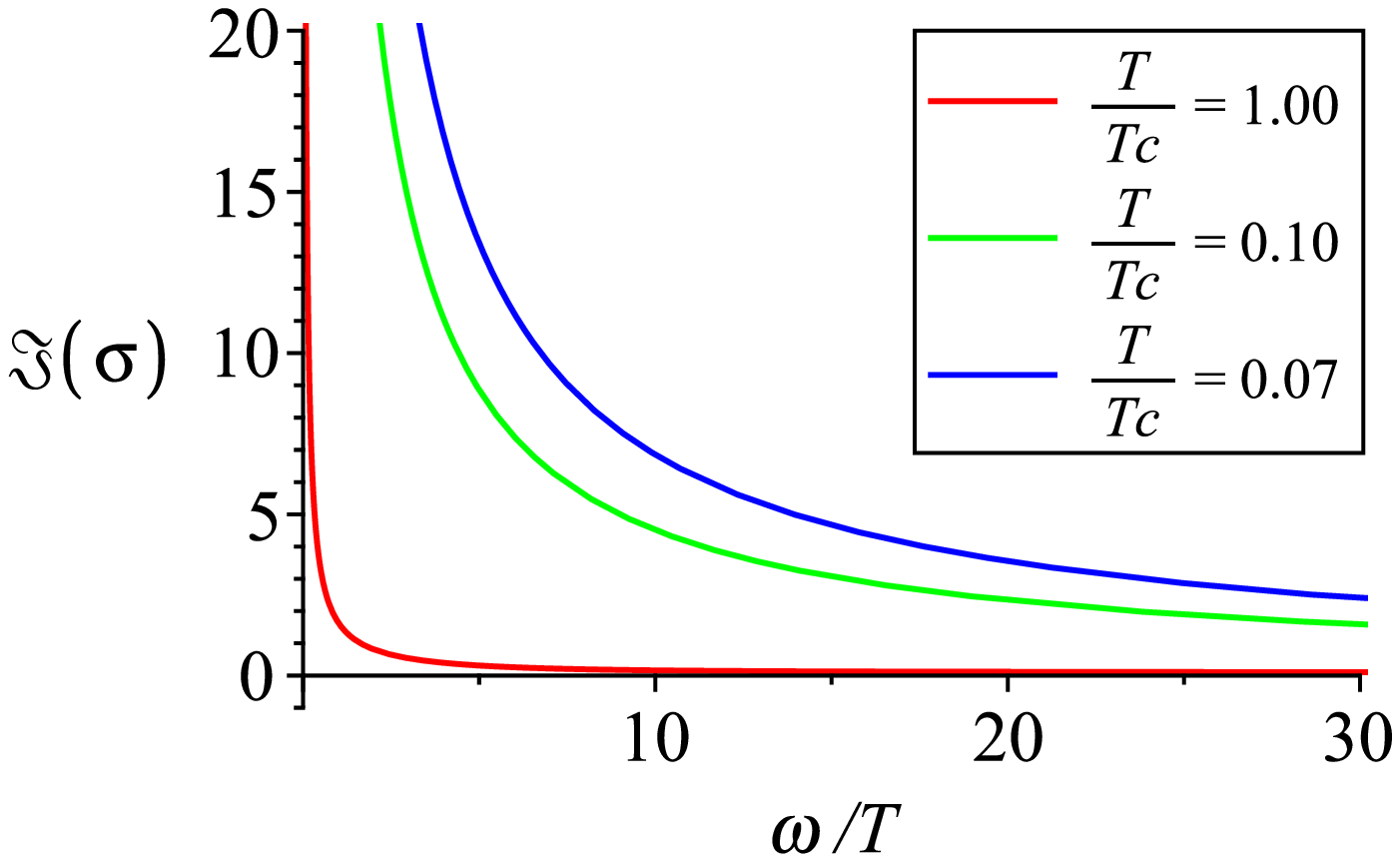}
\\ \hline
\end{tabular}
\end{table}
\caption{The real and imaginary part of the conductivity with $\alpha=1$ and $q=1,3,5$ from left to right.
The different curves in each figure are respect to different temperature ratio.}\label{conductivity full 1}%
\end{figure}

\begin{figure}[H]
\centering
\begin{table}[H]
\centering
\begin{tabular}[c]{ |c|c|c| } \hline
\centering
\includegraphics[width=0.3\textwidth]{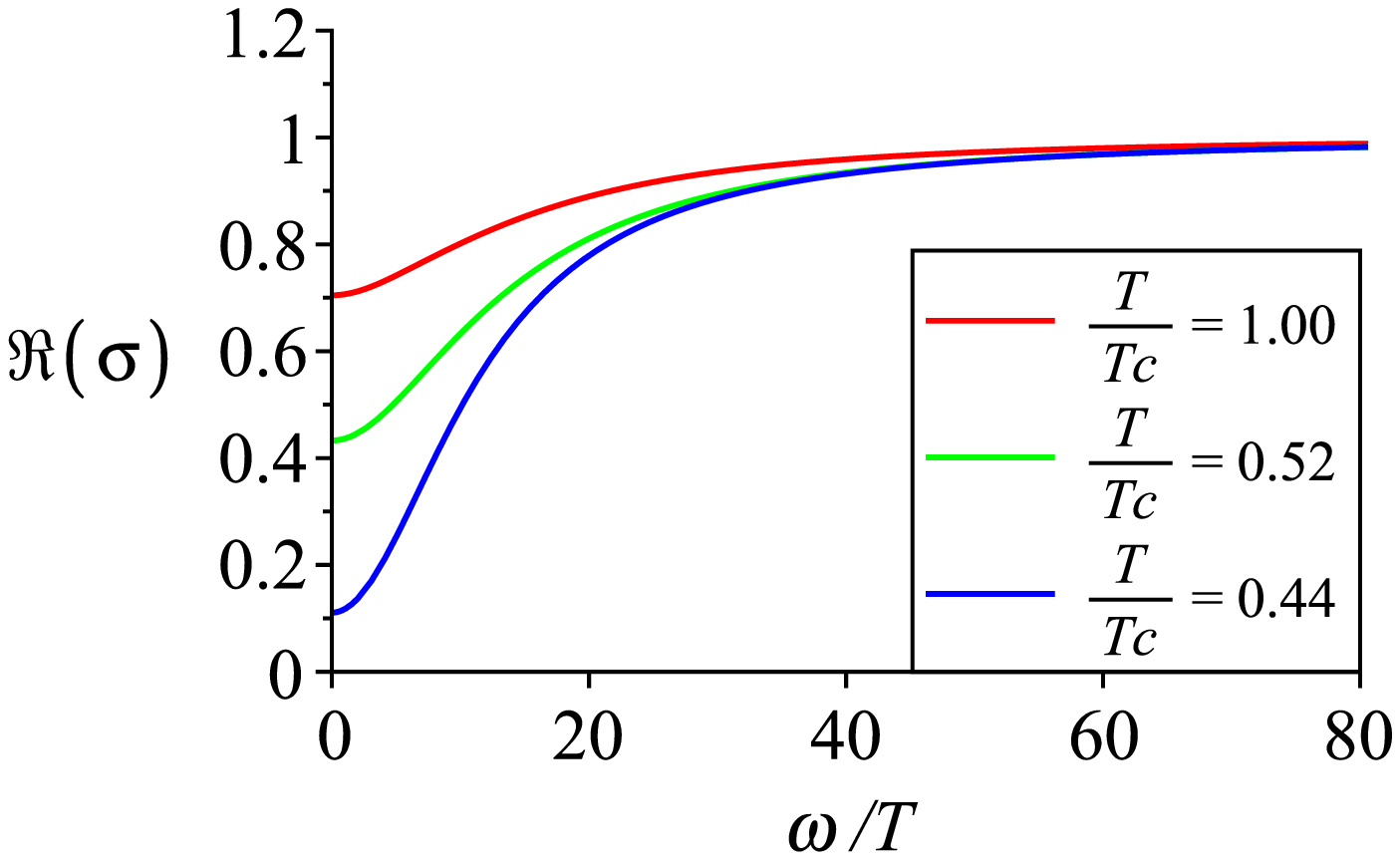}&
\includegraphics[width=0.3\textwidth]{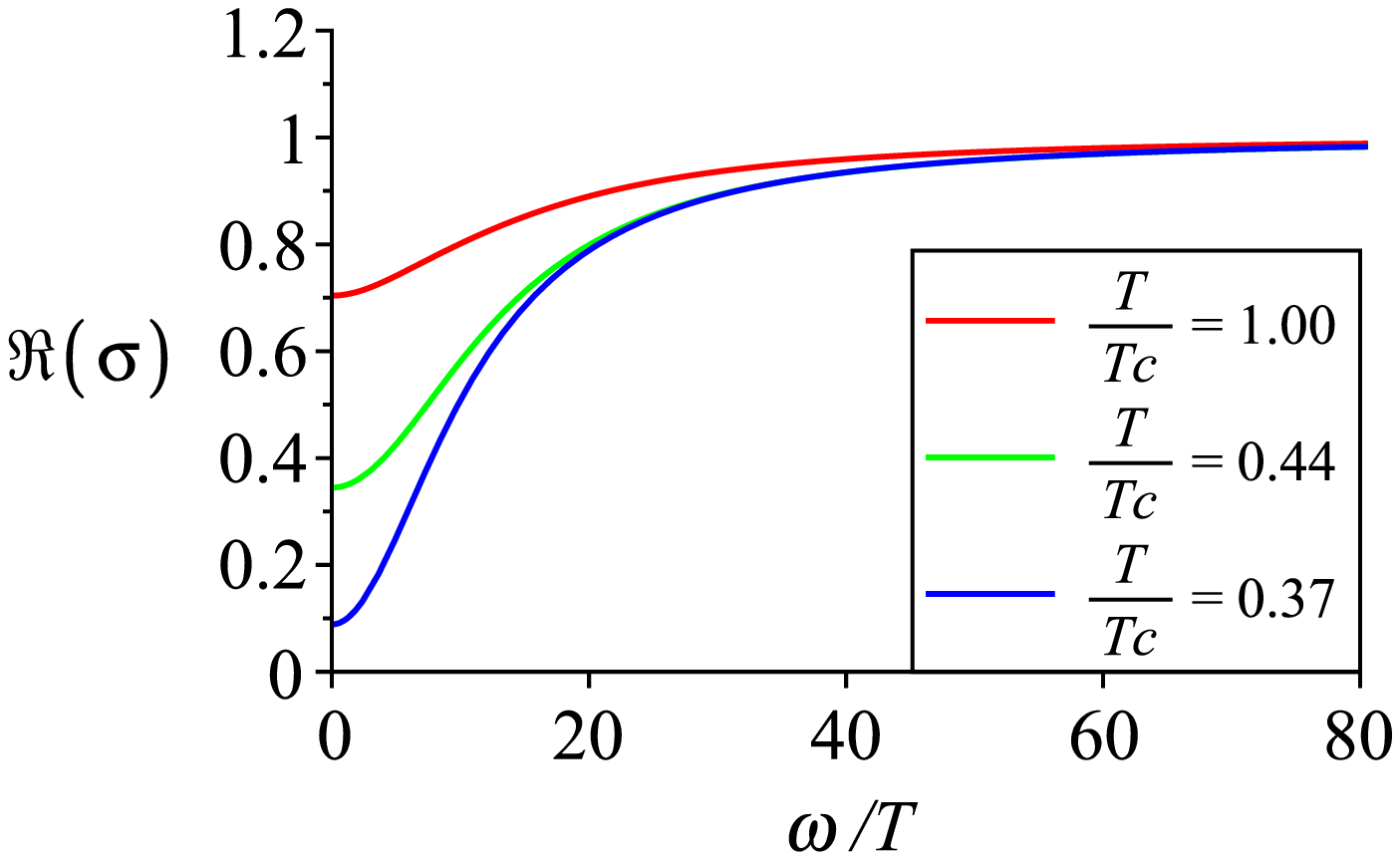}&
\includegraphics[width=0.3\textwidth]{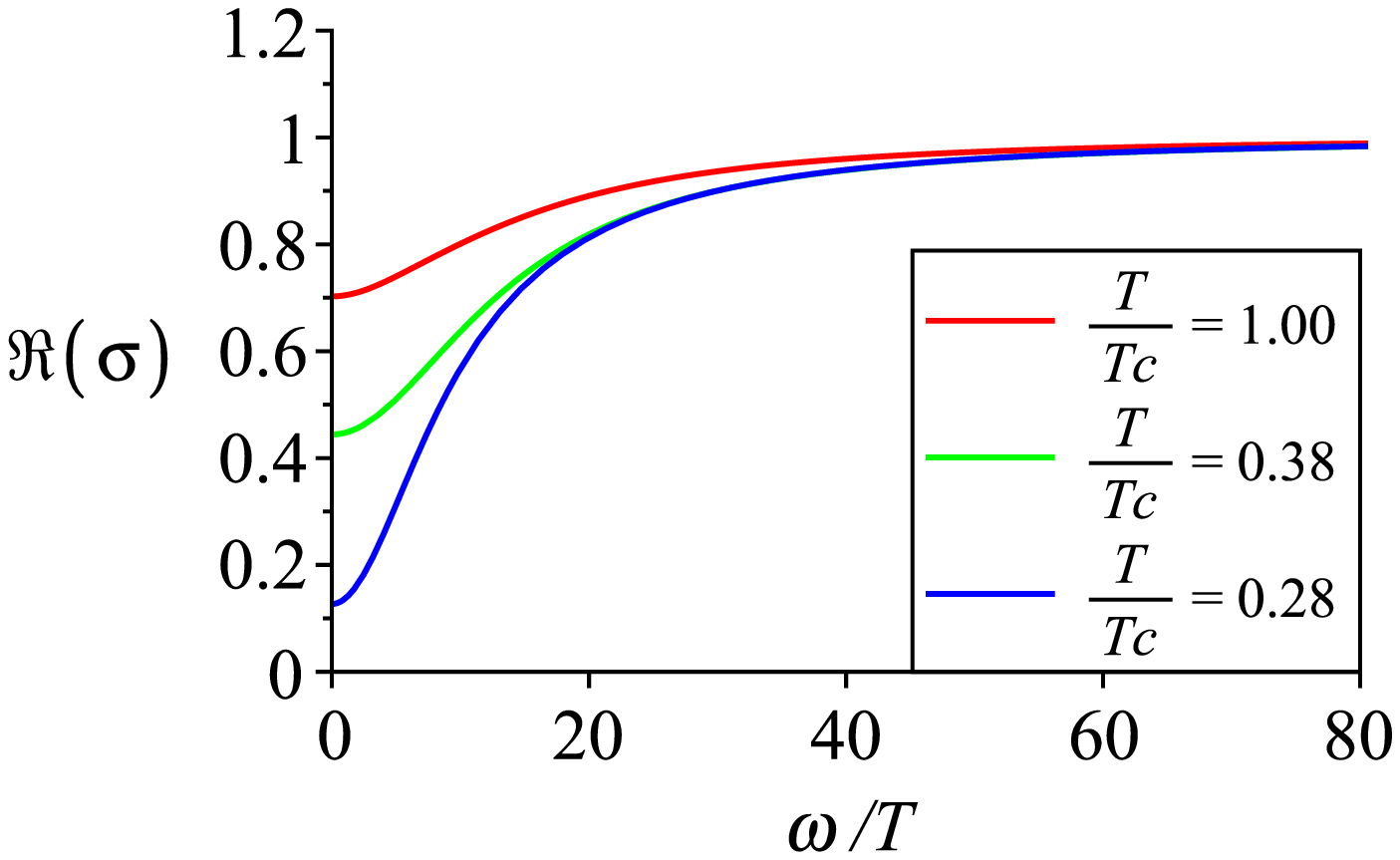}
\\ \hline
\includegraphics[width=0.3\textwidth]{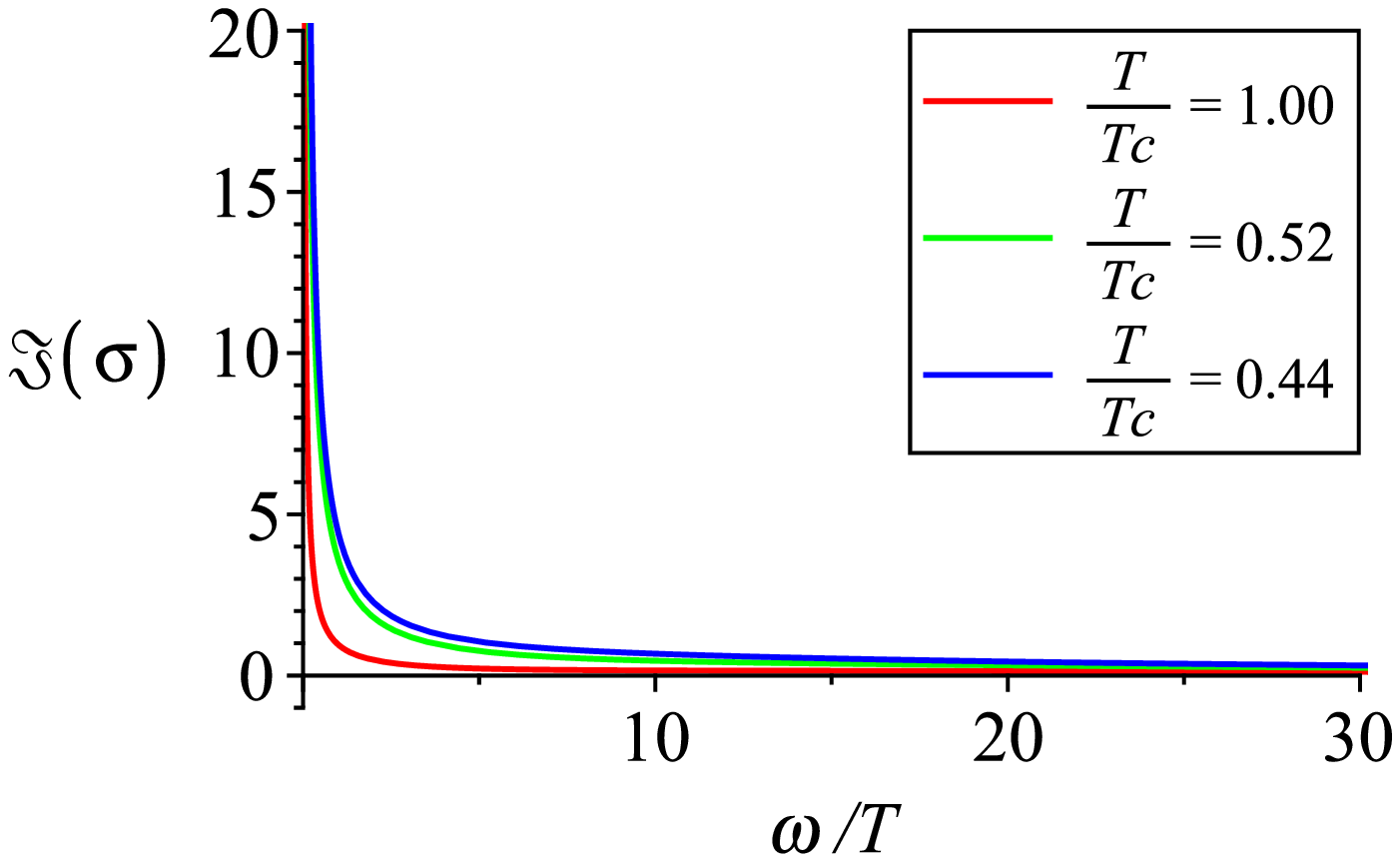}&
\includegraphics[width=0.3\textwidth]{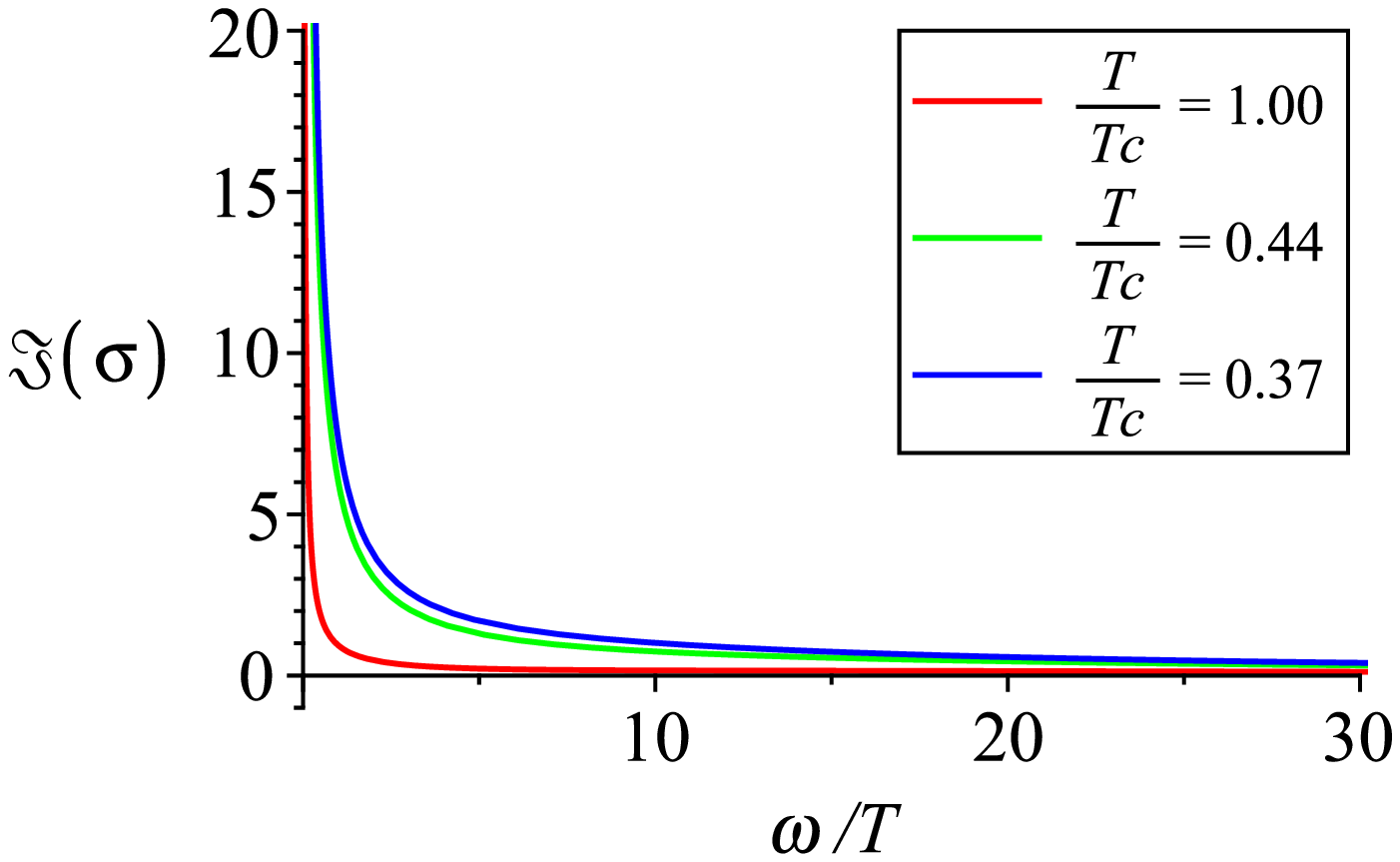}&
\includegraphics[width=0.3\textwidth]{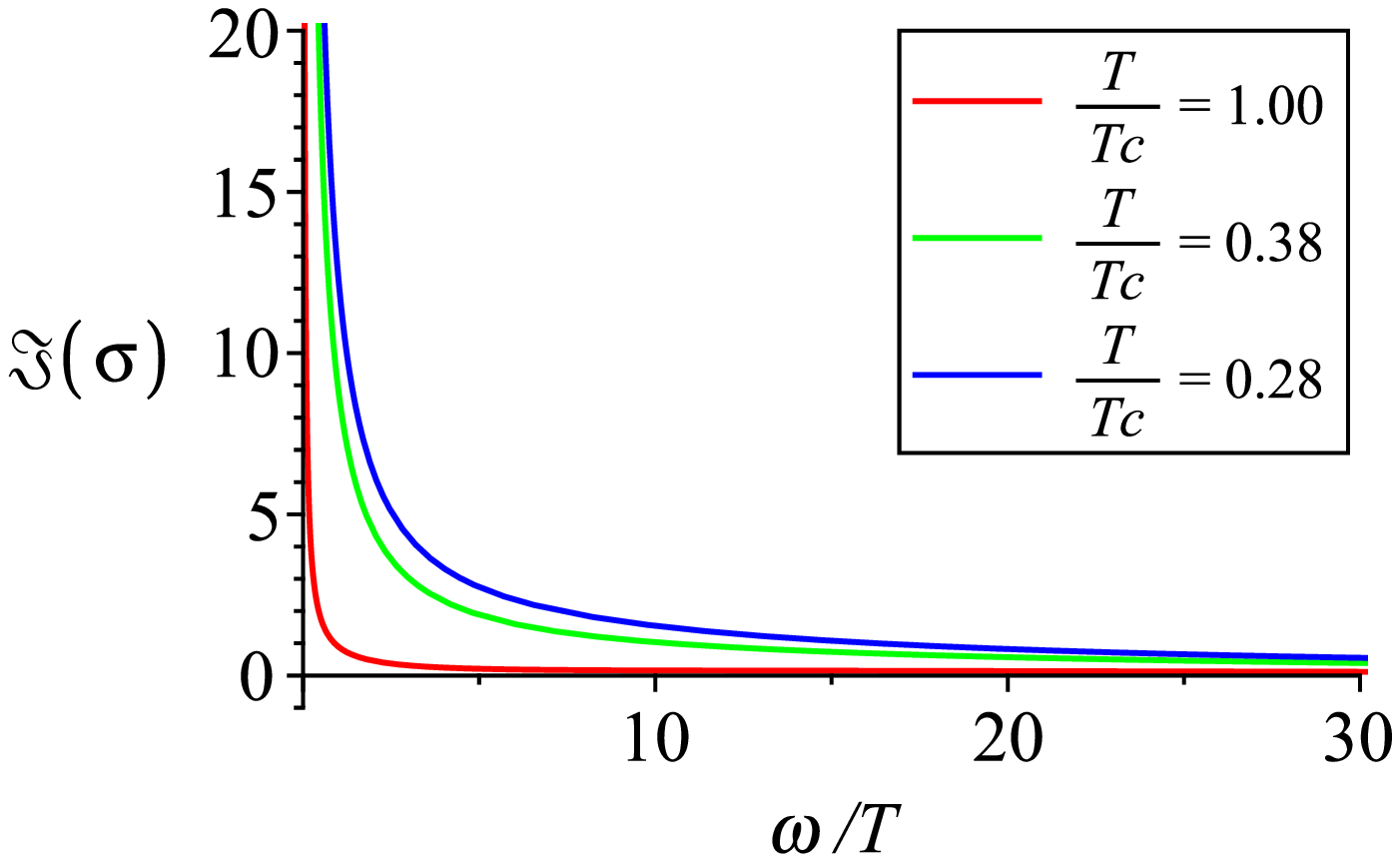}
\\ \hline
\end{tabular}
\end{table}
\caption{The real and imaginary part of the conductivity with $\alpha=3$ and $q=1,3,5$ from left to right.
The different curves in each figure are respect to different temperature ratio.}\label{conductivity full 2}%
\end{figure}

\begin{figure}[H]
\centering
\begin{table}[H]
\centering
\begin{tabular}[c]{ |c|c|c| } \hline
\centering
\includegraphics[width=0.3\textwidth]{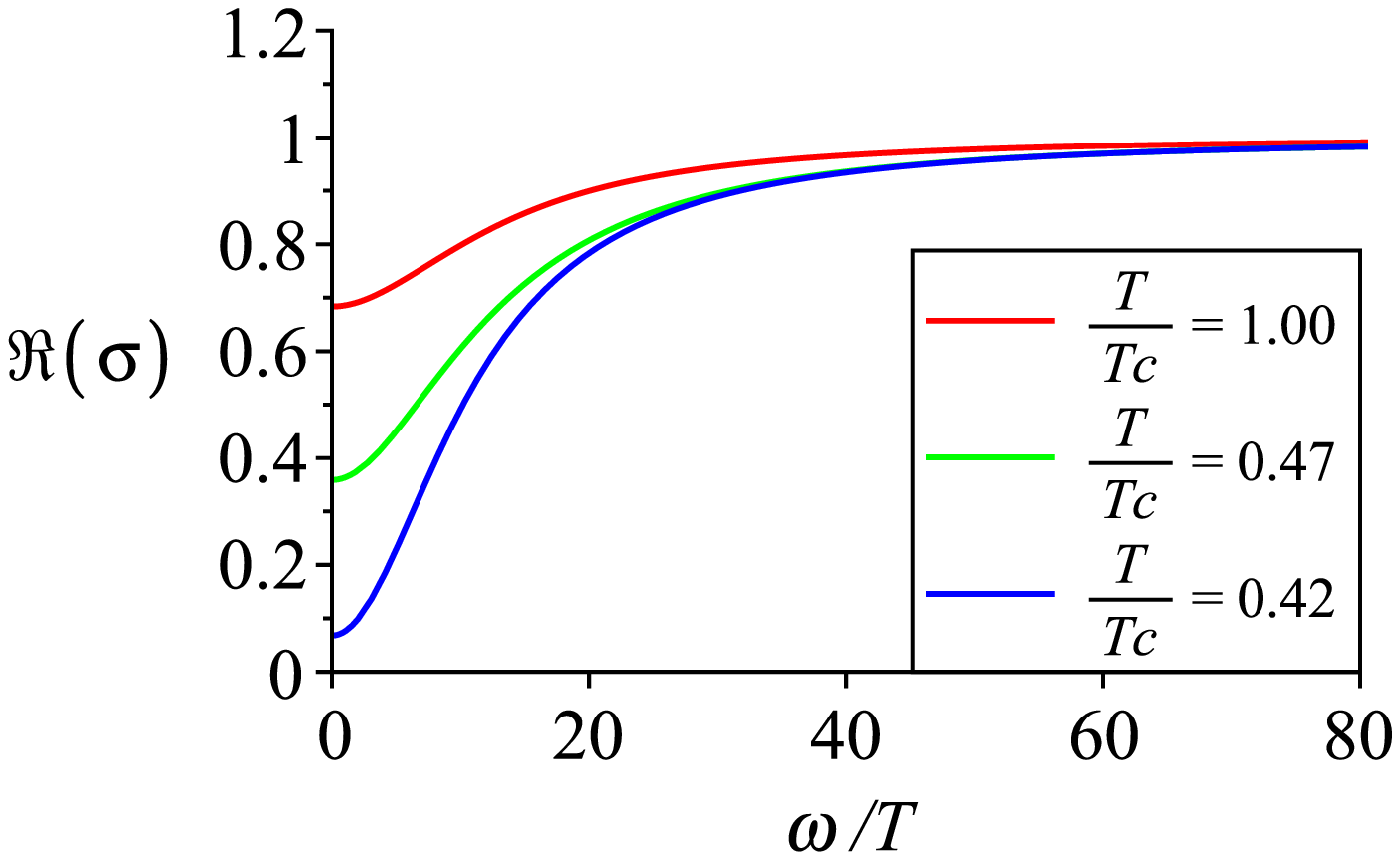}&
\includegraphics[width=0.3\textwidth]{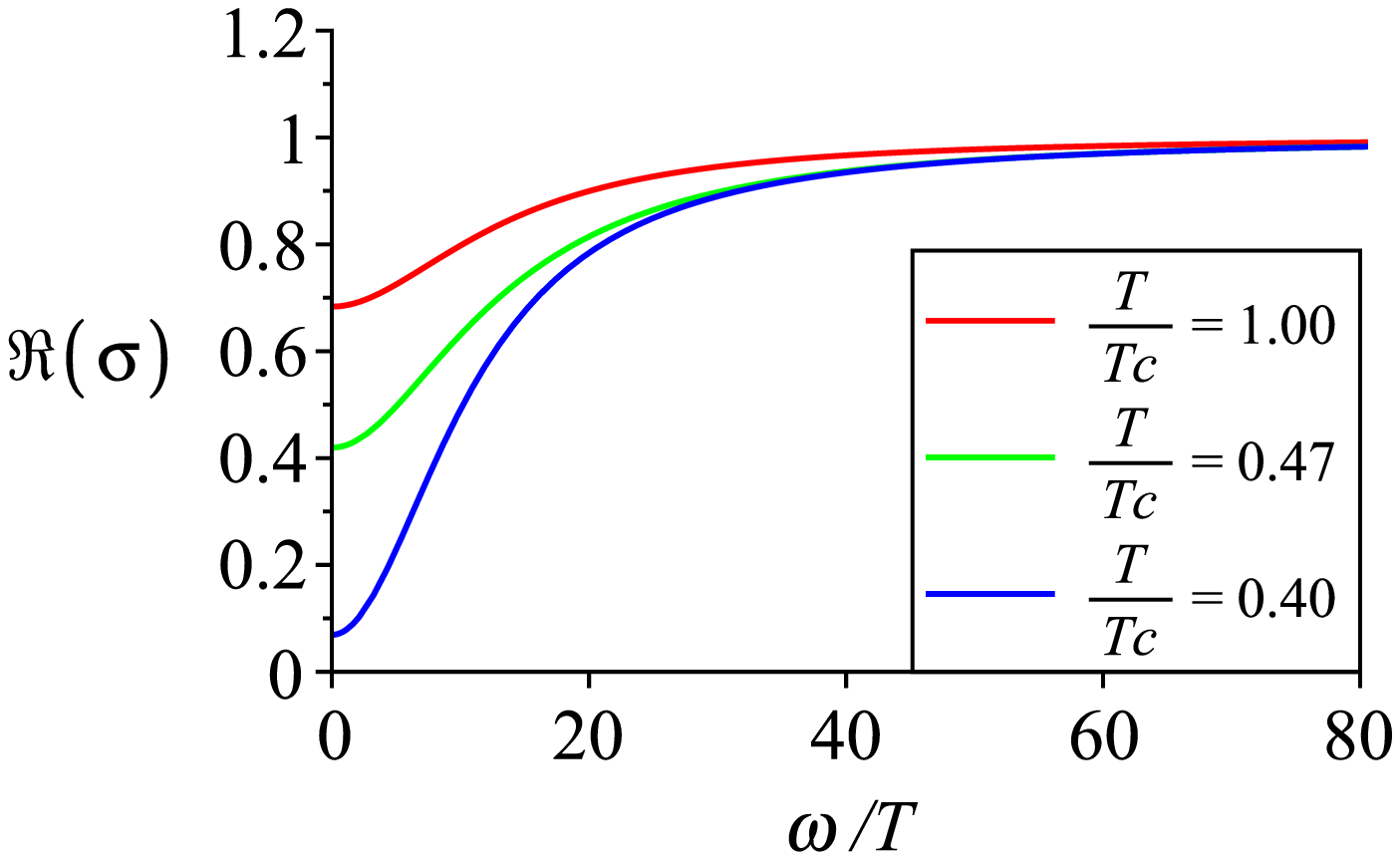}&
\includegraphics[width=0.3\textwidth]{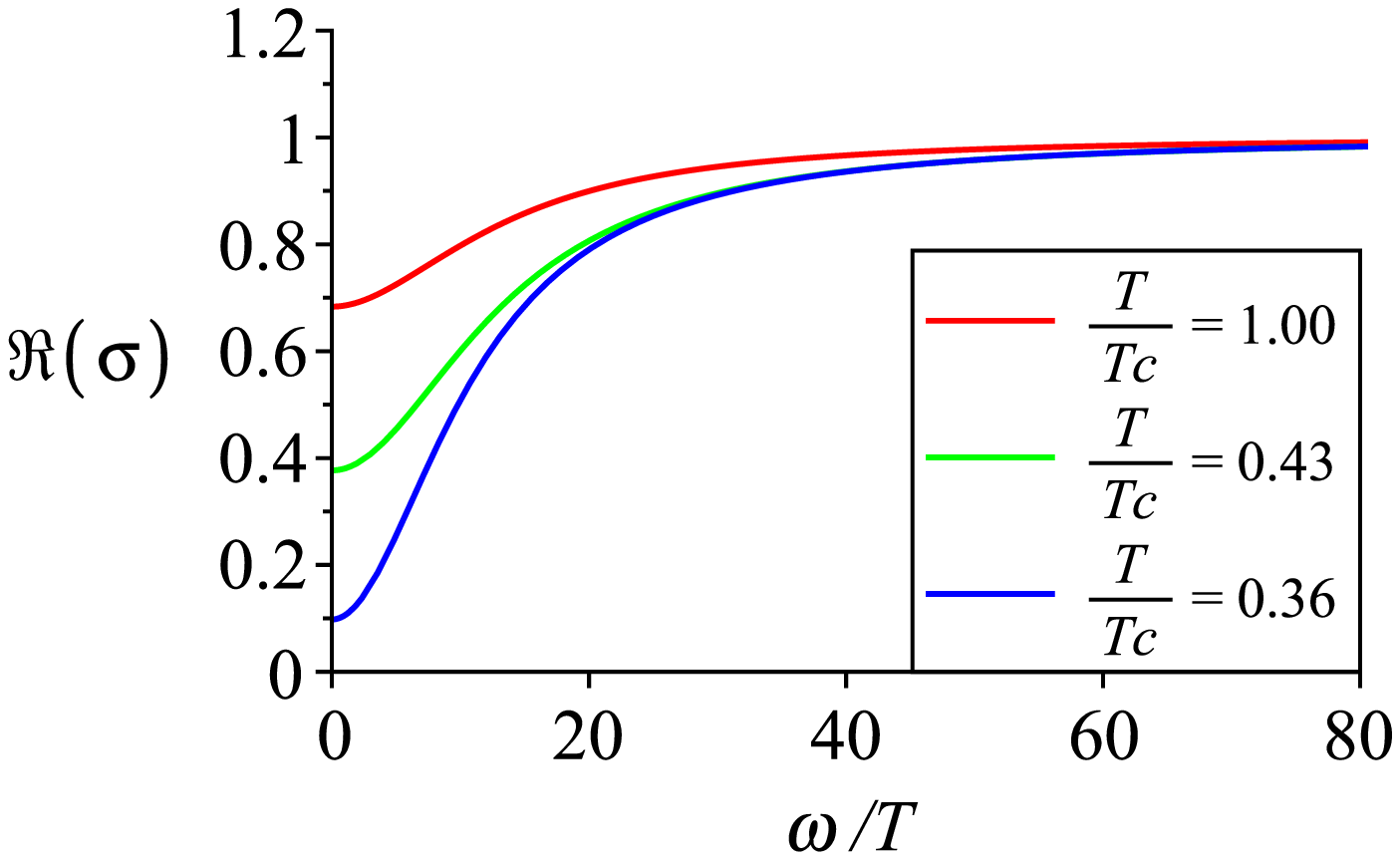}
\\ \hline
\includegraphics[width=0.3\textwidth]{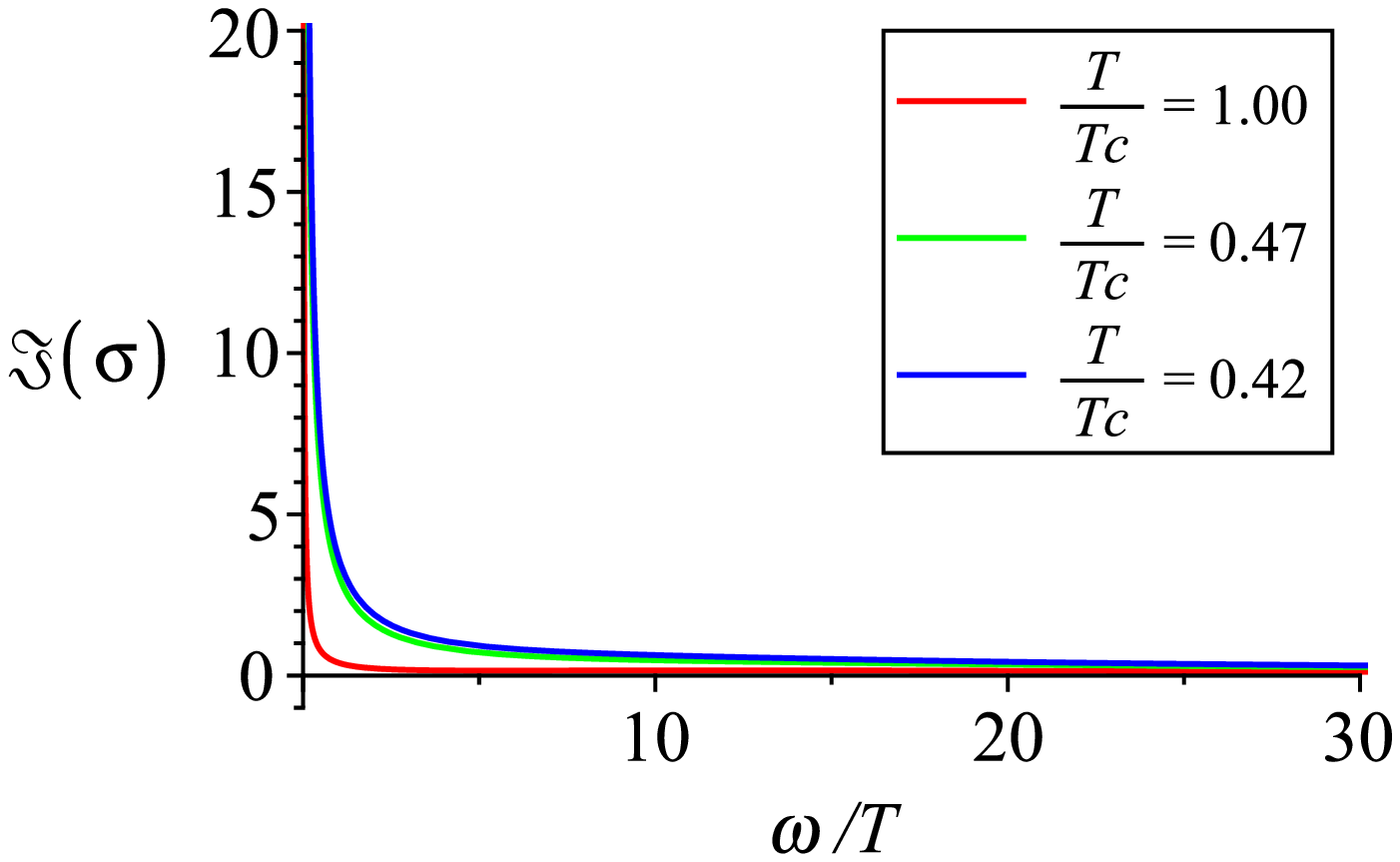}&
\includegraphics[width=0.3\textwidth]{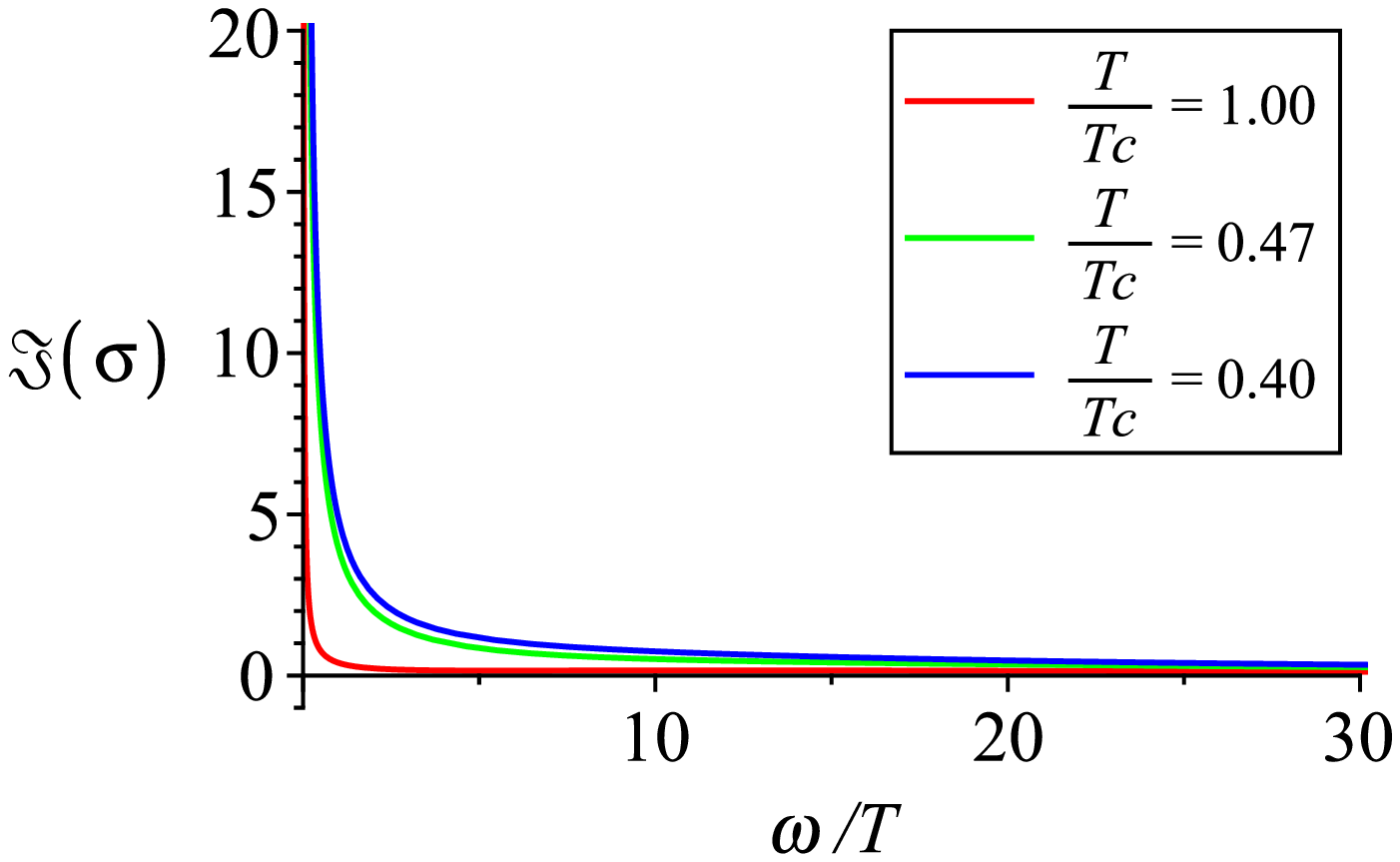}&
\includegraphics[width=0.3\textwidth]{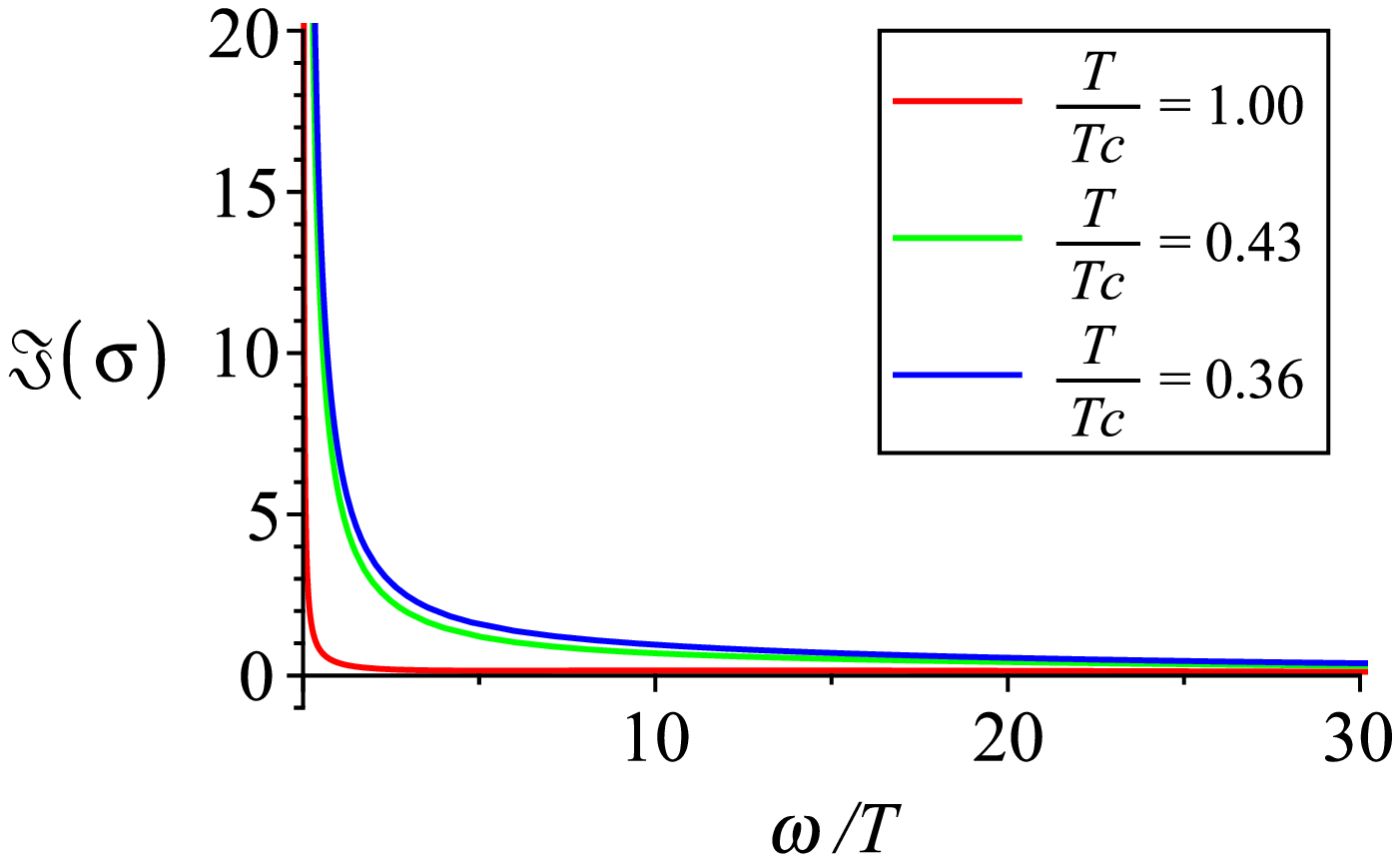}
\\ \hline
\end{tabular}
\end{table}
\caption{The real and imaginary part of the conductivity with $\alpha=5$ and $q=1,3,5$ from left to right.
The different curves in each figure are respect to different temperature ratio.}\label{conductivity full 3}%
\end{figure}

The real and imaginary parts of the conductivity vs. frequency are plotted in Fig. \ref{conductivity full 1} - \ref{conductivity full 3}. 

\setcounter{equation}{0} \renewcommand{\theequation}{\arabic{section}.\arabic{equation}}

\section{Conclusion}

We have seen that the generalized matching method can be regarded as collecting the boundary conditions which are located at several positions.
The well behaved fields can have a series expansion while all the degrees of freedom for the initial values are determined via the boundary conditions at the point of the series expansion.
It is not surprising that the simple boundary-conditions-collecting approach does work because of the achievement of the ordinary matching method.

In this paper, we studied a non-minimal HSC model by considering the Einstein-Maxwell-Dilaton system with both the non-backreaction and the full-backreaction cases by the matching method.
The model has two adjustable parameters $\alpha$ and $q$.
$\alpha$ describes the coupling between the Maxwell field and the dilaton, and $q$ represents the charge of the dilaton.

We found that the critical temperature $T_c$ of the phase transition does not depend on the backreaction of the gravitational background.
The critical temperature increases with  $\alpha$ and $q$ monotonously and behaves as $T_c\propto \alpha^{1/2}$ for small $q$ and  $T_c\propto q^{1/2}$ for small $\alpha$, which is consistent with the numeric results.
The condensate near the critical temperature behaves as $(1-T/T_c)^{1/2}$ which indicates that the phase transition is second order.

We showed that, in the non-backreaction case, the behavior of the condensate at low temperature blows up quickly and is not make sense, while in the full-backreaction case, the low temperature behavior of the condensate is rectified.
For a fixed $\alpha$, the condensate increases as the parameter $q$ increases, while for a fixed $q$, the condensate decreases as the parameter $\alpha$ increases.
The behavior of the condensate depending on the parameters $\alpha$ and $q$ is consistent with the numeric result.

In addition, we developed an approximate analytic method to calculate the electric conductivity.
We expand the perturbative field $A_x$ near the horizon obliged on the in-falling boundary condition of the Schrodinger equation at the horizon.
We then truncated the expansion to the linear order and determined the truncating coefficients by the equation of motion and the boundary condition at the boundary.
We analytically showed that the imaginary part of the conductivity suffers a $1/\omega$ divergence at small frequency that implies the Dirac $\delta$ function behavior in the real part of the conductivity, i.e. DC superconductivity.
Furthermore, we showed that the asymptotic values of the real conductivity in the small frequency limit increases as the temperature raised.
However, we did not observe the "Drude Peak" behavior at the low frequency as in the numeric calculation \cite{1006.2726}.

The matching method provides an analytic description of the HSC model. It help us to understand the analytic behaviors of the superconductors comparing to the numeric study. We believe that the analytic method could reveal more and more important properties of superconductors in the future.

\section*{Acknowledgements}
We thank Chiang-Mei Chen, Mei Huang, Ming-Fan Wu and Pei-Hung Yuan for useful discussion. This work is supported by the Ministry of Science and Technology (MOST 106-2112-M-009-005-MY3) and in part by National Center for Theoretical Science (NCTS), Taiwan.

\setcounter{equation}{0} \renewcommand{\theequation}{\Alph{section}.\arabic{equation}}

\begin{appendices}

\section{Condensate near critical temperature }

Expanding the condensate around the critical temperature $T_c$, we obtain the approximate condensate for $T\le T_c$,
\begin{subequations}
\begin{align}
D_{2}&=AT_c^2\left(-\frac{T-T_c}{T_c}\right)^{1/2},\\
A&=\frac{11264{\pi}^{4}{{T_c}}^{4}-81{\alpha}^{2}{\rho}^{2}}{5184{\pi}^{2}{{T_c}}^{4}}\left(-\frac{T_c}{T_1}\right)^{1/2},\\
T&=T_c+T_1 b^2+O(b^4),\\
T_1&=\frac{{\pi}^{4}\left(-2125824{\pi}^{4}{{T_c}}^{4}{\alpha}^{2}-77824{\pi}^{4}{{T_c}}^{4}{q}^{2}+22356{\alpha}^{4}{\rho}^{2}+5670{\alpha}^{2}{q}^{2}{\rho}^{2}+243{q}^{4}{\rho}^{2}\right){{T_c}}^{5}}{1944{\rho}^{2}\left({\pi}^{4}{{T_c}}^{4}{q}^{2}+10{\pi}^{4}{{T_c}}^{4}{\alpha}^{2}-{\frac{81{\alpha}^{4}{\rho}^{2}}{512}}\right)},
\end{align}\label{D_Tc_n}
\end{subequations}

\section{Formulas with full-backreaction }

The near horizon coefficients with full-backreaction are listed here.
The coefficients for the condensate background, (\ref{fHfull_}), are written as follows:
\begin{subequations}
\begin{align}
    \chi_{1}  &  = \frac{ -8\Big[ 16b^2{r_H}^4 + 4a^2(2\alpha^2b^2+1){r_H}^2 + a^4\alpha^4b^2 \Big] }{ \Big[ -4(b^2+6){r_H}^2 + a^2(\alpha^2b^2+2) \Big]^2 }, \\
    g_{1}  &  = \frac{ 4b^2{r_H}^2 - a^2(\alpha^2b^2+2) }{8{r_H}^2},\\
    B_{2}  &  = \frac{-6a}{ (\alpha^2b^2+2) \Big[ -4(b^2+6){r_H}^2 +a^2(\alpha^2b^2+2) \Big]^2 } \cdot \notag\\
    &  \cdot \Bigg\{  -16b^2\bigg[ \frac{1}{3} (\alpha^2+q^2)b^2 + 2\left(2\alpha^2+q^2-\frac{1}{3}\right) \bigg]{r_H}^4 \notag\\
    &  + \frac{4}{3}a^2\bigg[ \alpha^2(2\alpha^2+q^2)b^4 - (12\alpha^4-9\alpha^2-2q^2)b^2 + 2 \bigg]{r_H}^2 \notag\\
    &  + a^4\alpha^4(\alpha^2b^2 + 2)b^2  \Bigg\},\\
    \phi_{1}  &  = \frac{ 4b(4{r_H}^2 + a^2\alpha^2) }{ - 4(b^2 + 6){r_H}^2 + a^2(\alpha^2b^2 + 2) },\\
    \phi_{2}  &  = \frac{ -4b }{ (\alpha^2b^2+2) \Big[ - 4(b^2+6){r_H}^2 + a^2(\alpha^2b^2+2) \Big]^3  } \Bigg\{ - 64(b^2+12)(\alpha^2b^2+2){r_H}^6 \\
    &  + 16a^2\bigg[ \alpha^4(b^4+12b^2+6)b^2 - \alpha^2\bigg( (3q^2-2)b^4 + (18q^2-37)b^2 - 60 \bigg) - 2q^2b^2 - 12q^2 \notag\\&+ 10 \bigg]{r_H}^4- 8a^4\bigg[ \alpha^6(b^6+\frac{7b^4}{2}+9b^2) - \alpha^4\bigg( \left(\frac{3q^2}{2}-4\right)b^4 - \frac{27b^2}{2} + 6 \bigg) \notag\\&- \bigg( 4(q^2-1)b^2 - 21 \bigg)\alpha^2 - 2q^2 \bigg]{r_H}^2 + a^6\alpha^2(\alpha^2b^2+2) \cdot \bigg( (\alpha^2b^2+2)^2 + 2\alpha^2(2\alpha^2b^2-1) \bigg)
    \Bigg\}.
\end{align}\label{fBhc}
\end{subequations}

The coefficients of the near horizon expansion of the perturbed field $C(z)$ are:
\begin{subequations}
\label{Chc}
\begin{align}
  p_0 =&\frac{e^{\frac{\chi_0}{2}}}{({g_1}+3)r_H},\label{p0}\\
  C_{1} =& \frac{  \Gamma  }{  \Delta  }, \label{c1}
\end{align}
\end{subequations}
where,
\begin{align}
  \Gamma =&-2\Big({\alpha}^{2}(\chi_1-2){b}^{2}+2{\alpha}^{2}b\phi_1+2\chi_1-4\Big){\omega}^{2}{{e}^{\chi_0}}\notag\\&+4({g_1}+3)\Bigg\{\Bigg[\bigg[2{\alpha}^{2}\bigg(-{\omega}^{2}{{p_0}}^{2}-i\left(\frac{3}{8}\chi_1+\frac{1}{4}\right)\omega{p_0}+\frac{\chi_1}{8}\bigg){b}^{2}+i{\alpha}^{2}\omega{p_0}\phi_1(1-i\omega{p_0})b\notag\\&-4{\omega}^{2}{{p_0}}^{2}-i\left(\frac{3}{2}\chi_1+1\right)\omega{p_0}+\frac{\chi_1}{2}\bigg]{g_1}+\bigg[-9{\alpha}^{2}{\omega}^{2}{{p_0}}^{2}-i\left(\frac{9}{4}\chi_1+3\right){\alpha}^{2}\omega{p_0}\notag\\&+{q}^{2}+\frac{3}{4}{\alpha}^{2}\chi_1\bigg]{b}^{2}+3i{\alpha}^{2}\omega{p_0}\phi_1(1-i\omega{p_0})b+18\left(\frac{1}{3}-i\omega{p_0}\right)\bigg(\frac{\chi_1}{4}-i\omega{p_0}\bigg)\Bigg]{{r_H}}^{2}\notag\\&+i\omega({\alpha}^{2}{b}^{2}+2)\left(\frac{1}{2}-i\omega{p_0}\right)({g_1}+3){r_H}+\frac{1}{4}{a}^{2}({\alpha}^{2}{b}^{2}+2)^{2}\Bigg\},\\
  \Delta =&2({\alpha}^{2}{b}^{2}+2)\bigg({{e}^{\chi_0}}{\omega}^{2}+{{r_H}}^{2}({g_1}+3)^{2}(1-i\omega{p_0})^{2}\bigg).
\end{align}

The cubic equation of $a^2$ can be written as:
\begin{align}
  0 =& - 128(\alpha^2b^2+2) \cdot \bigg( b^6 + \frac{31b^4}{2} + 77b^2 + 114 \bigg){r_H}^6  + 96a^2\Bigg[ \alpha^4b^8 + (\frac{65\alpha^4}{6}+4\alpha^2)b^6 \notag\\&+ \bigg( 32\alpha^4 + \left(q^2+\frac{127}{3}\right)\alpha^2 + 4 \bigg)b^4 + \bigg(28\alpha^4 + \left(6q^2+\frac{335}{3}\right)\alpha^2  + \frac{2q^2}{3} + \frac{124}{3} \bigg)b^2 \notag\\&+ 40\alpha^2 + 4q^2 + \frac{302}{3} \Bigg]{r_H}^4  - 24a^4\Bigg[ \alpha^6b^8 + \left(\frac{37\alpha^6}{6}+6\alpha^4\right)b^6 + \bigg( \frac{23\alpha^6}{3} + (q^2+35)\alpha^4 \notag\\&+ 12\alpha^2 \bigg)b^4  - \bigg( 6\alpha^6 - 31\alpha^4 - \left(\frac{8q^2}{3}+66\right)\alpha^2 - 8 \bigg)b^2 + 4\alpha^4 + 26\alpha^2 + \frac{4q^2}{3} + \frac{124}{3} \Bigg]r_H^2 \notag\\& \cdot2a^6(\alpha^2b^2+2) \cdot \Bigg[ \alpha^6b^6 + 3\alpha^4\left(\frac{\alpha^2}{2}+2\right)b^4 - 2\alpha^2(2\alpha^4-3\alpha^2-6)b^2 + 2\alpha^2(\alpha^2+3) + 8 \Bigg]. 
\label{y3}
\end{align}

The DC expansion of the conductivity is:
\begin{align}
\label{sigDC}
  \sigma_{DC}\left(\omega\right) 
  =&i\sigma_{-1}\omega^{-1}+\sigma_{0}+O\left(\omega\right),\\
  =&i\left(1-\frac{2(\alpha^2b^2+2)(g_1+3)}{\Xi}r_H^2\right)r_H\omega^{-1}+\left(1-\frac{4p_0\left(\alpha^2b^2+2\right)\left(g_1+3\right)}{\Xi}r_H^3+\frac{\Phi}{\Xi^2}\right)\notag\\&+O\left(\omega\right),
\end{align}
where,
\begin{align}
  \Xi=&\left[\left(\alpha^2+\frac{4q^2}{\left(\chi_1+2\right)\left(g_1+3\right)}\right)b^2+2\right]\left(\chi_1+2\right)\left(g_1+3\right)r_H^2+a^2\left(\alpha^2b^2+2\right)^2,\\
  \Phi=&6\left\{p_0\Bigg[\left(\Big(\left(\chi_1+2\right)g_1+3\chi_1+8\Big)b-\frac{4}{3}\phi_1\left(g_1+3\right)\right)b\alpha^2+\left(2\chi_1+4\right)g_1+6\chi_1+16\Bigg]r_H\right.\notag\\&\left.-\frac{2}{3}\left(\alpha^2b^2+2\right)\left(g_1+3\right)\right\}\left(\alpha^2b^2+2\right)\left(g_1+3\right)r_H^4.
\end{align}

\end{appendices}

\bibliographystyle{unsrt}
\bibliography{HSC}

\end{document}